# First Observations With a GNSS Antenna to Radio Telescope Interferometer

J. Skeens[1] , J. York[1] , L. Petrov[2] , D. Munton[1] , K. Herrity[1] , R. Ji-Cathriner[1] , S. Bettadpur[3] , and T. Gaussiran[1]

[1]Applied Research Laboratories, The University of Texas at Austin, Austin, TX, USA, [2]NASA Goddard Space Flight Center, Greenbelt, MD, USA, [3]Center for Space Research, The University of Texas at Austin, Austin, TX, USA

**Abstract** We describe the design of a radio interferometer composed of a Global Navigation Satellite Systems (GNSS) antenna and a Very Long Baseline Interferometry radio telescope. Our eventual goal is to use this interferometer for geodetic applications including local tie measurements. The GNSS element of the interferometer uses a unique software-defined receiving system and modified commercial geodetic-quality GNSS antenna. We ran three observing sessions in 2022 between a 25 m radio telescope in Fort Davis, Texas (FD-VLBA), a transportable GNSS antenna placed within 100 m, and a GNSS antenna placed at a distance of about 9 km. We have detected a strong interferometric response with a Signal-to-Noise Ratio (SNR) of over 1,000 from Global Positioning System and Galileo satellites. We also observed natural radio sources including Galactic supernova remnants and Active Galactic Nuclei located as far as one gigaparsec, thus extending the range of sources that can be referenced to a GNSS antenna by 18 orders of magnitude. These detections represent the first observations made with a GNSS antenna to radio telescope interferometer. We have developed a novel technique based on a Precise Point Positioning solution of the recorded GNSS signal that allows us to extend integration time at 1.5 GHz to at least 20 min without any noticeable SNR degradation when a rubidium frequency standard is used.

**Plain Language Summary** We have developed a unique version of a device called a radio interferometer. This device usually uses two large radio telescopes to record data about the Earth or astronomical sources. Our device uses a Global Navigation Satellite Systems (GNSS) antenna and a Very Long Baseline Interferometry radio telescope instead of two radio telescopes. We hope to use it to link the two measurement systems. We tested it by using it to record signals from satellites and other astronomical objects. We found that it worked well and could detect signals from much further away than a GNSS antenna can detect alone. We also developed a new technique that allows us to continuously integrate signals for longer periods with a less precise clock without losing signal strength. This is the first time anyone has used a GNSS antenna with a radio telescope in this way.

## 1. Introduction

The method of Very Long Baseline Interferometry (VLBI), first proposed by Matveenko et al. (1965), has a number of applications in radio astronomy, astrometry, space navigation, and geodesy. The microwave frequency emission from a common source is received at multiple radio telescopes that digitize voltage readings, usually with 1- or 2-bits per sample and time stamps supplied by ultra-precise clocks. The first demonstrations of interferometric fringes with VLBI were published by Broten et al. (1967) and Brown et al. (1968). A comprehensive discussion of interferometry in astronomy can be found in Thompson et al. (2001). The elements of the VLBI network are commonly spaced hundreds or thousands of kilometers apart, and thus there is no physical connection between the sites. Radio telescope antenna sizes range from 6 to 500 m in diameter. For example, the Very Long Baseline Array (VLBA) antennas are 25 m in diameter (Napier & Peter, 1994). Natural radio sources detected with VLBI range from hundreds of kJy to tens of μJy, where 1 Jy = $10^{-26}$ W m$^{-2}$ Hz$^{-1}$. AGNs emit in a continuum with power-log spectra from tens of MHz to hundreds of GHz in frequency. Recordings of emission taken at different sites are correlated to produce a time series of auto and cross-correlation spectra. These spectra can be used to produce images or to determine station positions, source coordinates, and the Earth orientation parameters. In contrast to natural radio sources, GNSS satellites emit signals with flux densities of up to 6 MJy in bandwidths of about 20 MHz, and small patch antennas are typically used to record this narrowband emission.

Observational astronomy has historically advanced toward larger and more sensitive instruments. In this study, however, we proceed in the opposite direction and present a radio interferometer consisting of a VLBI antenna and a commonly used geodetic GNSS antenna with an effective diameter on the order of 10 cm. Because the interferometer sensitivity is proportional to the geometric mean of the System Equivalent Flux Density (SEFD) of its elements, through pairing a small GNSS antenna with a sensitive radio telescope, we will still be able to detect both strong emission from satellites and weak emission from natural radio sources.

This GNSS-radio telescope interferometer has a number of important applications in uniting the traditionally independent techniques of GNSS and VLBI processing. First, it can be used to produce vector tie measurements between the phase centers of its elements according to well established software and techniques. At short baselines of 30–5,000 m, phase delay ambiguities can always be resolved. Processing phase delays, a local tie vector between microwave antenna reference points can be measured with sub-millimeter repeatabilities, as was demonstrated by Rogers et al. (1978), Carter et al. (1980), Herring (1992), Hase and Petrov (1999), Varenius et al. (2021), and Niell et al. (2021). These vector ties are useful in several capacities, including as a repeatable link between a permanent GNSS monument and a VLBI radio telescope that may improve the realization of terrestrial reference frames (Ray & Altamimi, 2005). We anticipate that we will reach a similar repeatability with this interferometer, but this will involve calibrating many effects, including antenna phase variation, cable stretching and thermal effects, and gravitational deformation of the VLBI antennas.

As discussed in more detail in Petrov et al. (2023), vector ties generated with VLBI directly relate the microwave phase reference points. This direct measurement overcomes the difficulty in conventional optical surveying that models must relate the surveyed positions to the true electrical phase center with some associated errors. The ties could thus improve upon the accuracy and repeatability of total station or GPS kinematic local tie measurements as performed in Matsumoto et al. (2022) and Ning et al. (2015). Second, a GNSS receiver with a precisely known tie vector to a radio telescope that is an element of the global geodetic network can provide a measurement of a post-seismic motion. This measurement can be translated to the motion of the radio telescope within hours to avoid a degradation of the accuracy of Earth orientation parameters determined with VLBI (Dieck et al., 2023).

Additionally, the interferometer could provide a method of multipath calibration for monumented GNSS antennas, as the directional VLBI antenna is far more resistant to multipath than the omnidirectional GNSS antenna. Further, simultaneous collection of GNSS observables and VLBI data allows for the calculation of a clock solution with respect to the GPS time, which may permit time transfer to VLBI stations.

In this paper, we will detail the technology that made this interferometer possible, the software used in data processing, and the concept of its operation. We performed three experiments for validation and testing, and we will detail the methodology for producing clock solutions and applying clock corrections before using the collected data to show some preliminary demonstrations of the interferometer's sensitivity and performance. Future work will explore the geodetic implications of its observables and their use to the scientific community.

## 2. The Concept of the Interferometer

Figure 1 shows a simplified conceptual view of the GNSS-radio telescope interferometer, in which a VLBI radio telescope and a GNSS antenna observe a common celestial source or GNSS satellite and utilize VLBI processing to extract observables.

We selected short baselines for these demonstration experiments for several reasons. The AGNs, radio galaxies, and supernova remnants with the highest flux density in the L-band are spatially extended. Short baselines have coarse angular resolution and thus allow us to detect extended sources that would be resolved in long baseline observations. By using short baselines, we also eliminate the need for a dual band combination to remove the effect of the ionosphere, as radio waves traveling to both elements of the interferometer have common atmospheric delays. However, visibility data at short baselines are more susceptible to terrestrial Radio-Frequency Interference (RFI), so we had to take additional care in detecting signals from interfering terrestrial sources in the collected data and flagging affected visibilities.

Data analysis for GNSS satellites uses a different delay model due to the non-planar wavefront of the signals. Several previous works have included VLBI observations of GNSS satellites and developed near-field delay models for correlation and analysis, such as in Plank et al. (2017), Tornatore et al. (2014), and Hellerschmied et al. (2016).

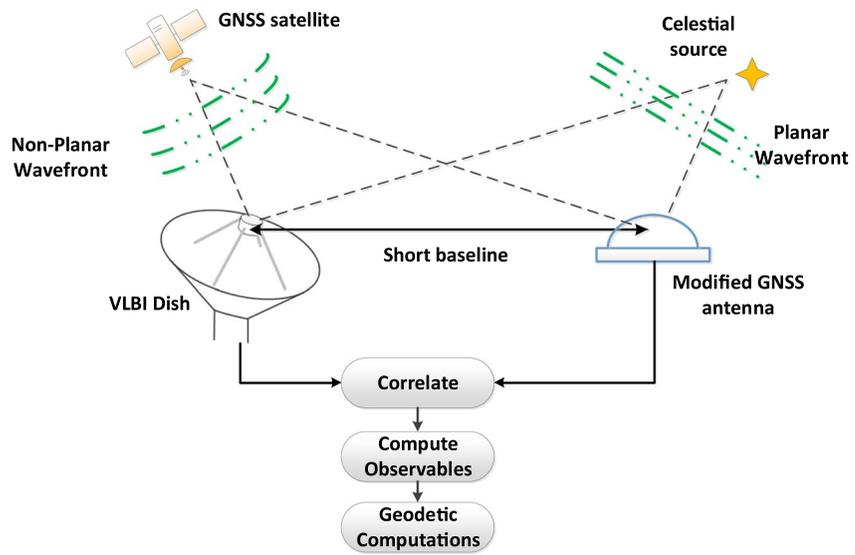

**Figure 1.** The conceptual design of the Global Navigation Satellite Systems (GNSS)-radio telescope interferometer. The celestial sources (right) are sufficiently distant that the incoming signals can be considered plane waves. However, depending on the baseline, the signals from the GNSS satellites may be in the near field and this must be taken into account. At the bottom of the diagram, we show a processing chain that reflects the standard Very Long Baseline Interferometry approach.

Although we plan to implement the satellite near-field delay model given in Jaron and Nothnagel (2019) for these and future experiments, no such model has been applied to the data shown here. To present reasonable fringe phase and amplitude measurements, we keep the coherent accumulation times for GNSS sources short.

Radio telescopes have a narrow beam pattern and with rare exceptions are fully steerable. They track sources during observations to compensate for Earth rotation. In contrast, a GNSS antenna has only limited directivity in the direction of local zenith and does not move during an observing campaign. This results in more susceptibility to RFI and a sensitivity that is strongly dependent on elevation angle.

VLBI systems also usually use extremely stable hydrogen masers that allow for the long integration times needed to detect faint sources. We were temporarily unable to access the hydrogen maser at the Ft. Davis VLBA site due to logistical issues including COVID restrictions, so we explored using relatively inexpensive rubidium frequency standards in the experiments detailed in this paper. These frequency standards have lower stability, and we must account for the expected clock variations to successfully integrate long enough for a detection on celestial sources. We developed a clock correction methodology to remove these variations that we will describe in detail.

## 3. The GNSS-Radio Telescope Interferometer

### 3.1. The GNSS Element of the Interferometer

The GNSS antennas we used are commercial, geodetic-quality Topcon CR-G5 GNSS antennas. The antenna receives right-hand circularly polarized emission, matching the polarization state of the GNSS signals. Off the shelf, the antenna electronics included hardware passband filters restricting signals to standard GNSS bands. These filters are implemented on a circuit board, making adjustment impractical by hand. To record emission in a wider bandwidth than the original antenna passband filters allowed, we replaced the circuit board installed on the antenna with connectorized components that is, components that can be bought individually and attached together through a mechanical interface. The GNSS antenna filter and amplification stage before and after the modifications is shown and labeled in Figure 2. The connectorized components consist of two high pass filters and two low noise amplifiers, with power supplied by a bias tee. All components are standard, commercially available products. The antenna element itself is unmodified, so we expect that these component modifications will have very little effect on the phase center of the antenna.

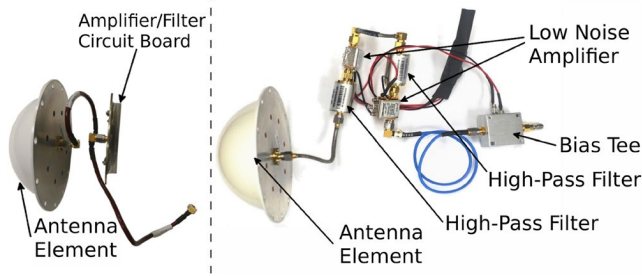

**Figure 2.** (Left) The original Global Navigation Satellite Systems antenna and circuit board amplification and filtering stage. (Right) The same antenna element with connectorized components used to expand the admitted passband.

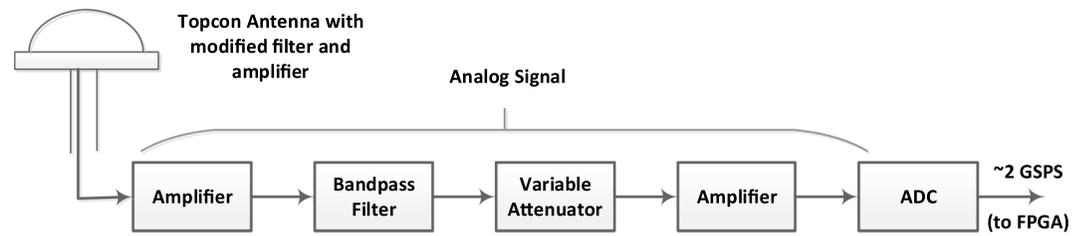

**Figure 3.** The path of analog High Rate Tracking Receiver data from the modified Global Navigation Satellite Systems antenna.

To accomplish the task of digitizing the signals collected at the GNSS antenna elements, we used the High Rate Tracking Receiver (HRTR), previously described by York et al. (2011), York et al. (2012, 2014). The HRTR is a software-defined GNSS receiving system that provides direct sampling of the L-band at approximately 2 Giga-samples per Second (GSPS), followed by real-time direct digital down conversion inside of a Field Programmable Gate Array (FPGA). The HRTR produces real time GNSS observables in a hardware accelerated processing chain as is common for GNSS receiving systems and digital data in raw baseband samples for later processing. These raw baseband samples are analogous to the data recorded in VLBI experiments.

Figure 3 shows the path of the received signal beginning with the modified antenna and filter components. After the custom-made amplification and filtering stage, the signal passes through another amplifier before entering a bandpass filter that restricts the signal to about 1.1–1.65 GHz. It then passes into a variable attenuator, which we tune to prevent saturation. A last amplification stage is followed by analog to digital conversion at 2 GSPS.

Figure 4 shows the end-to-end noise response of the radio frequency system before and after the antenna and filter modifications. We collected these data in a laboratory test with an Agilent N8974A NFA series noise figure analyzer connected to an N4001A noise source. The noise response is characterized by the effective receive chain system temperature, $T_{RX}$, which is referenced to the input terminals of the first low noise amplifier. We disconnected the antenna element in this test and injected noise directly into the radio frequency system just before the analog to digital converter in the digitizing front end of the HRTR. The variable attenuator after the bandpass filter in Figure 3 usually attenuates the signal passing to the analog-to-digital converter by 16 dB, but for these tests, we increased the attenuation to 31 dB to avoid saturation in the last amplifier before analog-to-digital conversion.

The receive chain system temperature therefore does not include the effect of losses from the antenna element and many external effects, such as antenna mismatch and pattern effects, spillover effects, ambient non-excised RFI, atmospheric losses, and galactic background. The unmodified signal path shows the two frequency bandpasses allowed for the GPS L1 frequency at 1,575.42 MHz and the L2 and L5 frequencies at 1,226.60 and 1,176.45 MHz. After modification, a much wider frequency spectrum is opened at high gain and low noise temperature.

Following analog-to-digital conversion, the digital signal passes to an FPGA, which performs digital downconversion and filtering. Figure 5 shows the path of this digital signal through the digital baseband receiving system of the HRTR. The tunable digital filters provide up to nine separately configurable frequency bands with a fixed 40.912 MHz bandwidth. The signal at each band is sampled with one bit in-phase ($I$) and one bit in-quadrature ($Q$) quantization. Here the digital signal splits into two paths. The first is a hardware accelerated system of GNSS processing including a correlation engine and loop engine. Pseudorange and carrier phase measurements are recorded as part of standard HRTR operation, and this hardware accelerated processing produces these conventional range and phase measurements in real-time, capturing the observables for all civil GNSS signals. We convert these data to RINEX format. Output data rates are configurable, but for the work described here, we generated observables at a 1 Hz rate. The other path of the digital signal saves the baseband I and Q samples directly to a RAID of magnetic hard drives.

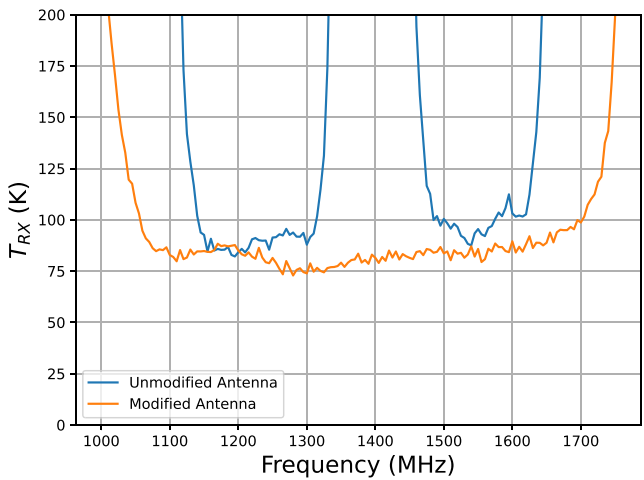

**Figure 4.** The receive chain effective system temperature by frequency before (blue) and after (orange) the filter and amplifier modifications in the Global Navigation Satellite Systems antenna.

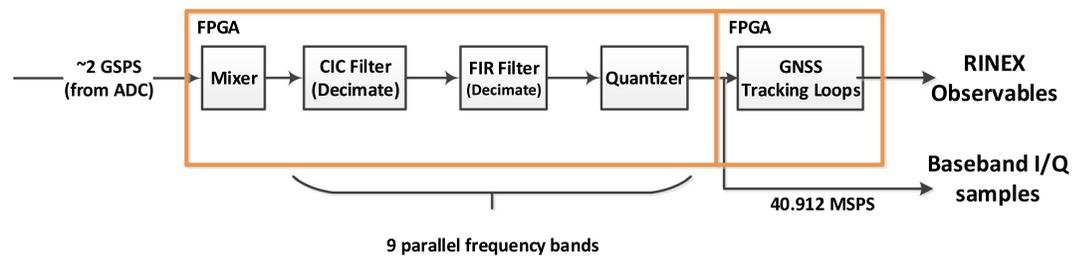

**Figure 5.** The path of digital High Rate Tracking Receiver data after analog to digital conversion at 2 Gigasamples per Second. Processing is hardware accelerated with multiple Field Programmable Gate Arrays to allow for real-time data flow, and the data is split into Global Navigation Satellite Systems observables and raw baseband *I/Q* samples.

The multiple selectable frequency bands provide the flexibility we need to observe both GNSS navigation signals and the bandwidth used in VLBA observations. Each frequency band has a maximum useable bandwidth of approximately 36 MHz of 40.912 MHz total bandwidth as a result of filter roll-off after the digital Cascaded Integrator-Comb (CIC) and Finite Impulse Response (FIR) filters. With nine configurable bands, this setup provides a maximum useable bandwidth of about 324 MHz.

The placement of additional frequency bands outside the GNSS bands is important for the detection of natural radio sources. The GNSS signals increase the in-band noise, which makes observations of celestial sources more challenging. In addition, there are terrestrial noise sources that interfere with observations, so it is useful to be able to identify and operate in quieter bands. More generally, wider bandwidth improves sensitivity and the precision of group delay measurements. During the experiments detailed here, we used between five and eight of the nine configurable frequency bands. We progressively added more simultaneously collected HRTR bands as we tested that all data transfer to disk could be handled in real time. One HRTR band was always placed at each of GPS L1 and L2 to collect observables for dual-frequency Precise Point Positioning (PPP). GPS L2 is outside of the bandwidth recorded by FD-VLBA. We thus used a bandwidth between 103 and 128 MHz in the relatively radio-quiet region of the VLBA L band coverage between 1,376 and 1,504 MHz sky frequency for natural radio source observations and a single HRTR band placed at GPS L1 for GNSS satellite observations. To produce high-precision geodetic data, we will likely need to calibrate small phase center location differences at each of the configurable bands in an anechoic chamber. This additional calibration is only necessary for natural radio source observations, as the GPS L1 and L2 phase centers are already calibrated for the Topcon CR-G5.

The GNSS observables and the *I/Q* baseband samples are stored in Hierarchical Data Format, version 5 (HDF5) files. We convert these data to VLBI Data Interchange Format (VDIF) for use in existing VLBI processing tool chains.

The HRTR requires an external stabilized 10 MHz frequency source as a time reference, and for these experiments we used a signal from either a GPS-disciplined rubidium frequency standard, or, at McDonald Geodetic Observatory (MGO), a hydrogen maser. The stability of the rubidium frequency standard is insufficient to support coherent accumulations over long (5–20 min) intervals without significant degradation of SNR due to decorrelation. Dealing with this degradation became one of our main technical challenges in analyzing data from celestial sources, and we will describe a post-processing technique for longer coherent accumulation intervals.

## 3.2. The VLBA Element of the Interferometer

The VLBA, as described in Napier and Peter ([1994](#)), consists of 25 m diameter parabolic antennas. The antennas utilize cryogenically cooled receiving systems, and ultra-stable hydrogen maser frequency standards provide the time references for the observing systems. The Fort Davis VLBA radio telescope used in this experiment has an L-band system temperature of 29 K and Degrees per Flux Unit (DPFU) 0.111 K/Jy for right-hand circularly polarized radio waves of wavelength 21 cm (1.427 GHz in frequency). This changes to 28 K and 0.105 K/Jy respectively at 18 cm wavelength (1.666 GHz). Measurements of sensitivity at both bands are distributed by

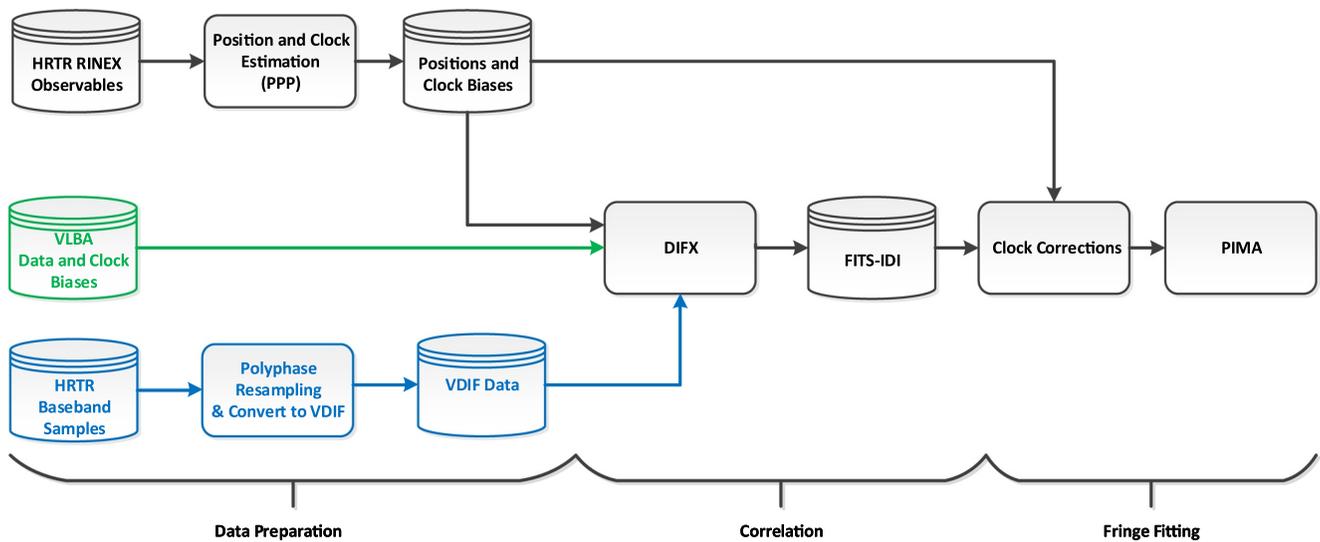

**Figure 6.** The software processing pipeline used in analyzing the collected data in all three experiments. The upper branch (in black) shows the data preparation for the Global Navigation Satellite Systems observables via a dual frequency Precise Point Positioning algorithm, which produces required position and clock estimates for the High Rate Tracking Receiver (HRTR) systems. In the middle, in green, the Fort Davis Very Long Baseline Array data is used in the Distributed FX correlation. At the bottom, the HRTR baseband *I/Q* samples undergo processing to adjust sampling rate and bit depth and are then converted to VDIF format and used in the correlation step. The results of the correlation step are delivered to PIMA to perform fringe fitting and produce observables.

National Radio Astronomy Observatories (NRAO) with periodic updates and are based on regular calibrated single dish experiments.

We used VLBA's digital downconverter observing mode with two single polarization bands of 128 MHz bandwidth. These data were sampled at 256 Megasamples per Second (MSPS) with real, 2-bit quantization and provided in VDIF. NRAO also provided time-tagged clock offsets with respect to GPS time for the hydrogen maser at the Fort Davis site.

## 4. Software Processing Pipeline

We performed processing with tools widely used by the VLBI community. There are two main components to this processing: correlation and fringe fitting. In the correlation step, signals from two antennas are Fourier transformed, multiplied with a signal offset determined by an a priori path delay, and summed in time and frequency. This produces complex visibilities at discrete frequencies and epochs. For celestial sources, we used an integration time of 1 s for correlation, and for GNSS sources, we used an integration time of 0.1 s. Although the complex visibilities are produced with an a priori model designed to remove Earth rotation and other effects, there are always residual delays and fringe rates present after correlation. The next step in the process, fringe fitting, applies more precise models to remove these effects in a parameter search of group delay, phase delay, and phase delay rate. Fringe fitting also coherently sums complex visibilities, allowing for far fainter signals to be detected.

Our approach for processing the data we collected with the HRTR and the VLBA radio telescope is shown in Figure 6. There are three main steps in this processing pipeline.

1. Pre-processing the baseband data.
2. Correlation with the Distributed FX (DiFX) correlator (A. Deller et al., 2007; A. T. Deller et al., 2011).
3. Fringe fitting with the software package PIMA (Petrov et al., 2011).

To execute the correlation step, we need to know the positions of our HRTR antennas and to obtain an estimate of the local clock biases. The rubidium frequency standard we used experiences drift over our measurement intervals, so evaluation of local clock behavior is necessary. For this step we use the PPP software package diffproc (Olson & Tolman, 2018), which ingests the HRTR dual frequency GPS carrier phase and pseudorange observations along with International GNSS Service (IGS) precise ephemerides to estimate the position of the antenna along with ionospheric path delays and local clock behavior. For these estimates, we used the full observation set collected during our observations at each HRTR site, which was about 3–4.5 hr in duration. The PPP solution's Kalman filter residuals converged to about the centimeter level. This was sufficient for these initial experiments.

To make the HRTR data suitable for correlation, we use a polyphase resampler to upsample the data to a compatible sampling rate of 64 MSPS to simplify the correlator setup. Bandwidth in radio astronomy is traditionally a power-of-two multiple of 1 MHz. The HRTR was designed for GNSS applications, and therefore implements a bandwidth and sampling rate that is a multiple of 10.228 MHz, which is a commonly used base frequency for GPS receivers. Fort Davis, Texas (FD)-VLBA used a recorded bandwidth of 128 MHz, which is not commensurate with the HRTR recorded bandwidth of 40.912 MHz. At the time of writing, the DiFX correlator has not been fully tested for data streams with bandwidths that are not a power-of-2 multiple of 1 MHz. We attempted to correlate the FD-VLBA signal against the HRTR signal without adjustment, but we got unsatisfactory results: the SNR was much lower than expected, and the spectra of the cross-correlation function exhibited a pattern of corruption. Instead of attempting to upgrade the source code of DiFX to accommodate hardware that has not yet been tested, we decided to implement an extra pre-processing step. That step involved resampling the HRTR signal to 64 MHz bandwidth. The polyphase resampling process involves upsampling by an integer factor, filtering with a zero-phase low pass FIR filter, and subsequently downsampling by an integer factor. In this case, the data are upsampled by a factor of 4,000 and downsampled by a factor of 2,557.

The resampling tool is a direct implementation in SciPy (Virtanen et al., 2020) of an algorithm described in Vaidyanathan (1993). The default parameters are used in the filtering process, meaning that values beyond the filter boundary are assumed to be zero, and a Kaiser window with shape parameter $\beta = 5$ is used in the low pass FIR filter. After resampling, the data is re-quantized using 2 bits for the in-phase and 2 bits for the quadrature samples, and the quantization level threshold is set according to the optimal level for data with a low correlation coefficient (F. Schwab, 1986). The number of bits used to represent each sample is increased to 2 from 1 after resampling to mitigate SNR degradation caused by the resampling process.

We then convert the HRTR data to VDIF. This is a straightforward procedure in which the baseband $I/Q$ samples are extracted from HDF5 files along with relevant metadata, and the samples are subsequently rewritten in 32,000-byte data frames, each with a 32-byte header describing the seconds from the reference epoch, the frame length, the number of bits per sample, and the station and thread IDs. Unfortunately, VDIF does not provide room for coding metadata. Therefore, most metadata stored in the HRTR HDF5 files are communicated to DiFX through configuration files.

The next step in the processing pipeline is to compute the cross-correlation of the signals between the two antennas, and for this we use version 2.6.2 of the software package DiFX (A. T. Deller et al., 2011). Because of the nonstandard setup of the HRTR, we manually edited DiFX configuration files to input the custom frequency configurations in each of the three experiments. We convert the results of the correlation to a Flexible Image Transport Transport System (FITS) Interferometry Data Interchange Format (IDI) file (E. Greisen, 2022) using the difx2fits routine from the DiFX package.

After correlation, we use a novel processing technique to adjust the phases of the complex visibilities stored in the FITS file using the clock biases derived from the PPP solutions with a method described in Section 6. This allows longer coherent accumulations during the fringe fitting process, which is critical for observations of fainter celestial sources. The clock correction is essential for data collected with a rubidium frequency standard, but it may be beneficial even for data collected with a hydrogen maser frequency standard to allow for the removal of higher order clock terms in geodetic solutions. We fringe fit the data with release 9 January 2023 of the software package PIMA, which is described in detail in Petrov et al. (2011) and was written primarily for use in absolute astrometry and geodesy experiments.

## 5. Experiment Design

### 5.1. Physical Configuration

We conducted three experiments to test and verify the operation of the GNSS-VLBA interferometer in January, April, and September of 2022. The GNSS element of the interferometer consists of a modified GNSS antenna with a choke ring, a tripod, a rubidium clock, a HRTR, a general use laptop, a propane-powered electric generator, and coaxial cables. The two GNSS systems fit inside the trunk of a sport utility vehicle and setup took less than an hour. All three experiments included three collecting antennas—two GNSS antennas at MGO and Fort Davis, Texas (FD), named MGO-GNSS and FD-GNSS, and the VLBA antenna at FD, named FD-VLBA. This gave the interferometer two long baselines, FD-VLBA/MGO-GNSS and FD-GNSS/MGO-GNSS, each of about 9 km length and one short baseline, FD-VLBA/FD-GNSS of about 100 m length. Figure 7 shows the relative placement of these antennas with aerial photography courtesy of the U.S. Geological Survey. The MGO-GNSS antenna used a 10 MHz signal from a rubidium frequency standard in the first experiment conducted on 26 January 2022. For the following two experiments, the MGO-GNSS antenna was stabilized by the hydrogen maser at the station MACGO12M. Figure 8 shows a schematic of the setup in the 21 April and 28 September experiments, wherein MGO-GNSS used a hydrogen maser clock.

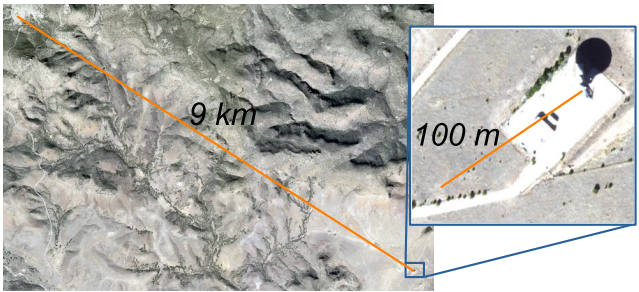

**Figure 7.** The interferometer configuration near Fort Davis, TX. The long baseline between the antenna at McDonald Geodetic Observatory and Fort Davis, Texas is shown on the left. The short baseline near the Fort Davis, Texas Very Long Baseline Array antenna is given in the enlarged section at the right.

## 5.2. Observation Configuration

Table 1 summarizes the configurations used in each experiment. We used the software package sur_sked for scheduling the experiments. During observing sessions, the antenna slews to the target source, tracks the source while recording baseband samples, and slews to another target. The period of time when a signal from a given source was recorded is called a scan. Scan lengths were typically fixed to 600 s for observing natural sources and either 40 or 60 s for observing satellites. The scheduling process for observing natural sources was the same as for VLBI astrometry observations: elevations and azimuths of targets were computed with a step of 1 min, targets were prioritized based on their elevations, and a sequence of slewing and tracking is formed such that the slew time between scans is minimized.

Scheduling GNSS satellites is more complicated because we must account for their motion during a scheduling segment. To do this, we used North American Aerospace Defense Command (NORAD) Two-Line Element (TLE) orbital elements to compute the time series of satellite Cartesian coordinates in an Earth fixed coordinate system with a time step of 1 s using an SGP4 propagator. We then calculated apparent topocentric right ascensions and declinations with a time step of 1 and 30 s. The latter time series was used for building a source table. An entry of the source table contains pointing positions with names derived from a satellite name followed by a time-dependent suffix, apparent topocentric right ascensions and declinations, and the mid-epoch of the validity time range (30 s). The scheduling preparation process ran in two passes. During the first pass, all source entries with mid-epochs within 30 s of the scheduling moment of time are considered. A score is assigned to each potential next target. The score depends on slewing time, elevation, and the history of observations of a given satellite. A preliminary schedule is generated by selecting the sources with the greatest score. A schedule contains right ascension, declination, scan duration, and start time for the observation of each target. However, the right ascensions and declinations were computed from a TLE with a coarse time resolution, and are therefore inaccurate. These errors are fixed in the second pass: right ascensions and declinations are re-computed to the middle epochs of the scans. The schedule is converted to key and vex format, which are used for steering VLBA antennas and for correlation respectively.

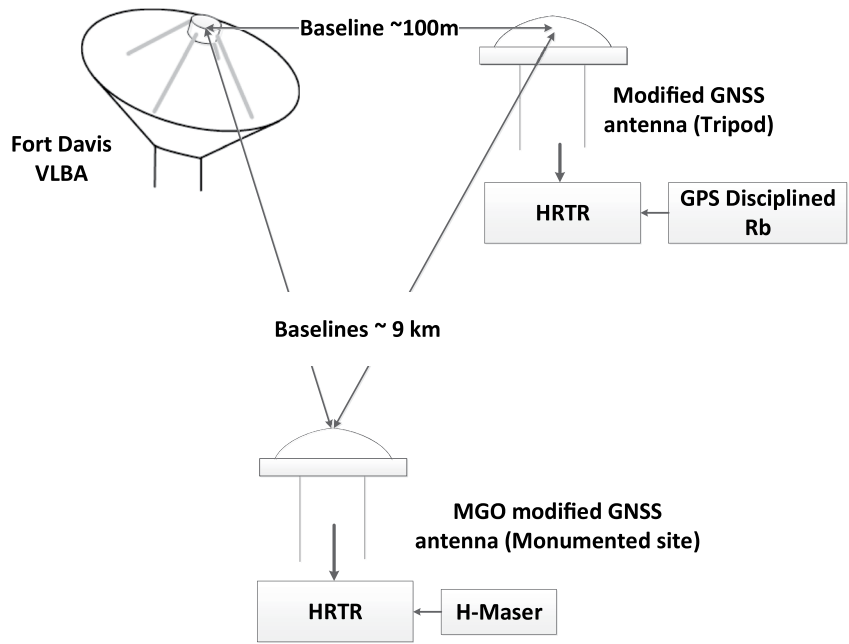

**Figure 8.** The physical design of the conducted experiments. Note that MGO-GNSS was deployed on a survey tripod, and it used a rubidium frequency standard in the 26 January 2022 experiment.

**Table 1**
*Dates, Frequency Standards, and Source Types in Each of the Three Experiments Referenced in This Paper*

| | Clock type | | Sources observed | | | |
|---|---|---|---|---|---|---|
| Experiment date | FD-GNSS | MGO-GNSS | GPS | Galileo | Celestial | OH maser |
| 26 January 2022 | Rubidium | Rubidium | Yes | No | Yes | No |
| 21 April 2022 | Rubidium | H Maser | Yes | No | Yes | No |
| 28 September 2022 | Rubidium | H Maser | Yes | Yes | Yes | Yes |

The accuracy of positions computed with SGP4 TLE propagation under these conditions is on the order of 1 km, which may cause an error in angular position up to 1:20,000, that is, 50 μrad or 10″. Since the full width half power of a VLBA antenna at 1.6 GHz is 27.5′ (Petrov, 2021), this pointing error is insignificant.

The a priori path delays required for correlation were computed based on apparent satellite right ascensions and declinations. These computations have significant inaccuracies because the satellites are located at distances of ~20,000 km, and the path delays are computed under the assumption that a source is located at infinity. The goal of these experiments was not precise geodesy and observations were made at short baselines with scans of up to only 60 s, so we compensated the inaccuracy in the a priori model by extending the fringe fitting search window. During these observations, we did not account for the motion of satellites during scans.

We observed three classes of objects. The first class is GNSS satellites. The second class is natural radio sources, AGNs, and supernova remnants. For these experiments, we selected celestial sources of the highest flux density. These sources tended to be significantly more extended than those typically used in geodetic applications, and many were therefore resolved on the longer FD-VLBA/MGO-GNSS baseline, leading to non-detections.

The third class is Galactic OH maser sources. We considered two major types of OH maser. The first originates in molecular clouds in star forming regions, and the second from circumstellar OH masers, which occur around large stars on the asymptotic giant branch of the Hertzsprung-Russell diagram and in red supergiant stars as they expel their outer layers in large, spherically symmetric envelopes (Cohen, 1989). These sources emit high flux density (up to 1,000 Jy) radio waves at specific frequencies corresponding to the emission lines of the hydroxyl radical. In this study, we attempted observations of the 1,612, 1,665, and 1,667 MHz emission lines for both molecular cloud and circumstellar masers. We picked the circumstellar masers from the largest flux densities available in the database developed in Engels and Bunzel (2015) along with one strong molecular cloud maser (W3(OH)). We only observed OH masers in the third experiment on 28 September 2022. For all experimental modes, the VLBA antenna tracked the source to account for the rotation of the Earth.

## 6. Analysis

### 6.1. Clock Corrections

For each of the three experiments, we collected L1 and L2 GPS observables starting 1 hr prior to our VLBA scheduled start until approximately 1 hr after the completion of the VLBA observations. We generated position and clock estimates using dual-frequency PPP fed with the GNSS observables and final orbit and satellite clock solutions from the IGS. The a posteriori standard deviations of the GNSS antenna position residuals from the PPP Kalman filter were between 3 and 10 cm. The median a posteriori standard deviation of the clock bias estimate residuals for each antenna and experiment were between 0.10 and 0.16 ns. All positions and uncertainties can be found in the Supporting Information S1.

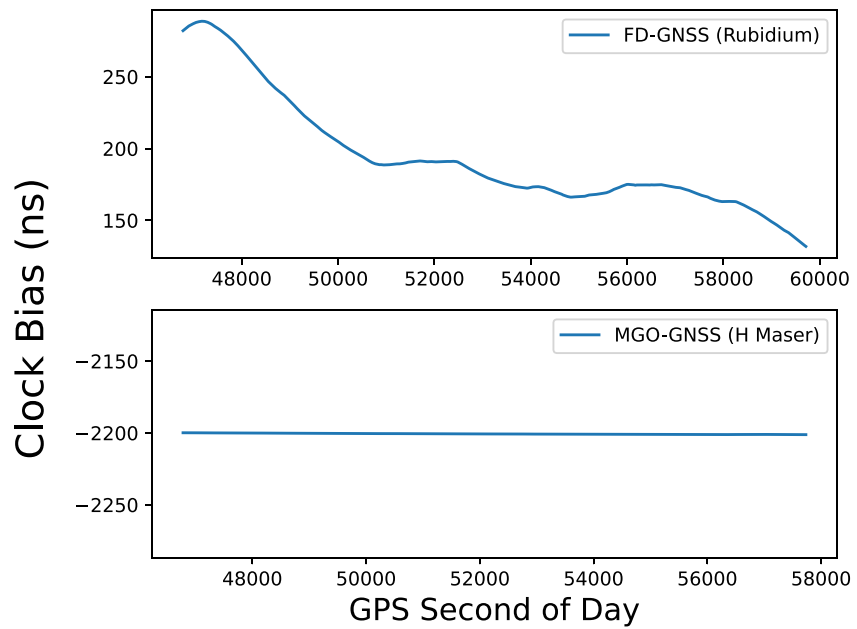

**Figure 9.** The Precise Point Positioning clock solutions for FD-GNSS (top, rubidium clock) and MGO-GNSS (bottom, hydrogen maser clock) for 21 April 2022.

We used the clock solutions computed from the GPS observables in two ways. First, we use the clock solution to obtain the clock offset at the beginning epoch of the observing session, which we used as an input to DiFX, slightly modifying the a priori delay model for both GNSS stations. Second, we used the detailed clock solutions to adjust the phase of the complex visibilities output by DiFX. The PPP software outputs the clock state at 30 s intervals, and we linearly interpolated between the clock states to find clock biases to compute phase corrections at the time tags of the complex visibilities. This phase correction was only applied to the GNSS stations that used a rubidium frequency standard. Figure 9 provides an example of the FD-GNSS and MGO-GNSS clock solutions in the April experiment. In the April experiment, the FD-GNSS system used a rubidium frequency standard, which explains the 100 ns variation over the observation interval. By comparison, the MGO-GNSS system was stabilized by a hydrogen maser, which explains the much smaller 1 ns of clock variation over the observation interval.

The relatively large nonlinear variation of the rubidium clock means that the maximum coherent accumulation interval for a given source is diminished due to the phase error introduced by the clock. By contrast, the hydrogen maser clock allows for long coherent accumulation periods without correction due to its stability. The clock solutions for the remaining experiments can be found in the Supporting Information S1 document.

We apply the clock solutions as a correction to the phase of the complex visibilities. To demonstrate, let $\theta(\nu, t)$ be the fringe phase correction in radians calculated from the PPP solution as a function of frequency and time. We denote phase in spectral channel $j$ and time epoch $k$ as $\theta(\nu = \nu_j, t = t_k) = \theta_{jk}$. We then define the clock bias in seconds corresponding to the phase correction as $\phi_k$, where $\phi_0$ is the clock correction at the beginning of the scan. The clock bias in seconds and the fringe phase correction are related for each spectral channel and epoch as,

$$\theta_{jk} = 2\pi\nu_j(\phi_k - \phi_0) \tag{1}$$

The clock correction is then applied to each complex visibility as a complex exponential phase rotation:

$$V_{jk}^* = V_{jk}e^{-i\theta_{jk}} = V_{jk}e^{-j2\pi\nu_j(\phi_k - \phi_0)} \tag{2}$$

With the PPP clock correction, we can coherently sum an entire scan, increasing the coherence time from about 200 s before the application of the clock solution to at least the full scan length of 1,200 s. To show that we have successfully lengthened the maximum coherence time, we can examine the SNR determined by PIMA based on the coherent accumulation interval given for fringe fitting. The SNR, $R_{sn}$, of a detection in the summation of the complex visibilities is given by Equation 6.45 of Thompson et al. (2001),

$$R_{sn} = \eta_q \sqrt{\frac{T_{ant,GNSS}T_{ant,VLBA}}{T_{sys,GNSS}T_{sys,VLBA}} 2\Delta\nu\tau_a}, \quad (3)$$

where $\tau_a$ is the coherent accumulation interval, $\Delta\nu$ the observation bandwidth, $\eta_q$ the quantization loss, and $T_{ant,GNSS}$, $T_{ant,VLBA}$, $T_{sys,GNSS}$, and $T_{sys,VLBA}$ are antenna and system temperatures. Figure 10 shows that the trend predicted in Equation 3 is satisfied after the clock correction is applied. The peak of the SNR in the uncorrected data occurs at roughly 200 s, after which the SNR begins to degrade. With the PPP clock solution applied, SNR continues to increase, and there is no sign of degradation after 20 min of coherent accumulation. For natural radio sources with lower flux densities, the additional coherent accumulation time permitted by this clock correction is critical for detecting the emission with an SNR high enough for geodetic analysis.

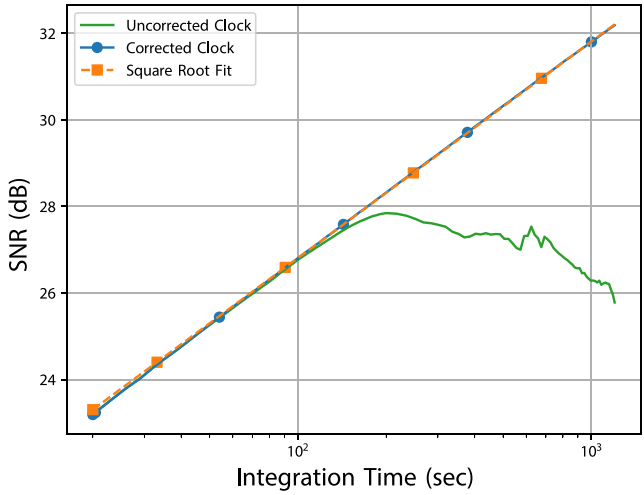

**Figure 10.** The Signal-to-Noise Ratio of a coherent accumulation of length integration time with and without the Precise Point Positioning clock solution for station FD-GNSS and source 3C 405 in the 21 April 2022 experiment.

### 6.2. Clock Correction Verification

To evaluate the PPP-derived clock correction against the phase variation in the correlated complex visibilities, we used short-duration coherent accumulations on a high-SNR source to derive an estimate for residual phase variation over time. We then compared that residual phase variation to the PPP-derived clock estimate. The governing equation for fine fringe fitting in the VLBI analysis software PIMA is shown below in Equation 4, where $w_{jk}$ is the weight applied to the complex visibility $V_{jk}$, and the fitted parameters $\tau_p$, $\tau_g$, $\dot{\tau}_p$ are the phase delay, group delay, and phase delay rate respectively.

$$C(\tau_p, \tau_g, \dot{\tau}_p) = \sum_j \sum_k V_{jk} w_{jk} e^{-2\pi i(\nu_0\tau_p + (\nu_j - \nu_0)\tau_g + \nu_0\dot{\tau}_p(t_k - t_0))} \quad (4)$$

The clock error contributes to the phase delay rate term of the fit, which captures linear variations in phase over time. We ran a number of trials and found that 200 s was an effective interval for which this linear fit would capture the clock variation of the GPS-disciplined rubidium clock within a wavelength, avoiding decorrelation and phase wrapping. To form the empirical fit, we ran the fringe fitting software for sequential 200 s intervals in PIMA. Thus, we obtain the "empirical" fringe phase correction,

$$\theta_k^{emp} = 2\pi(t_k - t_0)\nu_0\dot{\tau}_p + \theta_k^* \quad (5)$$

$\dot{\tau}_p$, the phase delay rate, is computed independently in each sequential 200 s fringe fitting along with group delay and phase delay terms. The time series correction $\theta_k^*$ is the remaining fringe phase time series after the fitted group delay, phase delay, and phase delay rate are removed through phase rotation:

$$\theta_k^* = \operatorname{atan}\left(\frac{\Im\left(C_k^{rot}\right)}{\Re\left(C_k^{rot}\right)}\right) \quad (6)$$

where $C_k^{rot}$ is the frequency-summed, fringe-fitted complex visibility time series:

$$C_k^{rot} = \sum_j V_{jk} w_{jk} e^{-2\pi i(\nu_0\tau_p + (\nu_j - \nu_0)\tau_g + \nu_0\dot{\tau}_p(t_k - t_0))} \quad (7)$$

We then coherently combine the empirical clock correction, $\theta_k^{emp}$, into a single time series covering the full scan for a given source. The 200-s sliding windows of fringe phase had one overlapping epoch with each neighboring sliding window, so to combine these series we shift the phase of the later 200-s interval such that it is equal to the phase of the previous 200-s interval at the overlapping epoch.

For the brightest and longest observed source in the three experiments—the 1,200-s Cygnus A (3C 405) scan in the 21 April 2022 data, Figure 11 shows the comparison of the PPP clock solution against the empirical clock solution. The two fringe phase corrections show very close agreement the fact that the clock correction derived from PPP and the PIMA-derived empirical phase trajectory show the same trends suggests that Because the empirical clock solution is the residual phase from a strong radio source after expected effects due to geometry are removed, the fact that the clock correction derived from PPP and the PIMA-derived empirical phase trajectory show the same trends suggests that clock drift in the rubidium frequency standard is the dominant error source after the a priori delay model is applied in DiFX. These results indicate that the PPP clock solution is sufficient for effective clock correction.

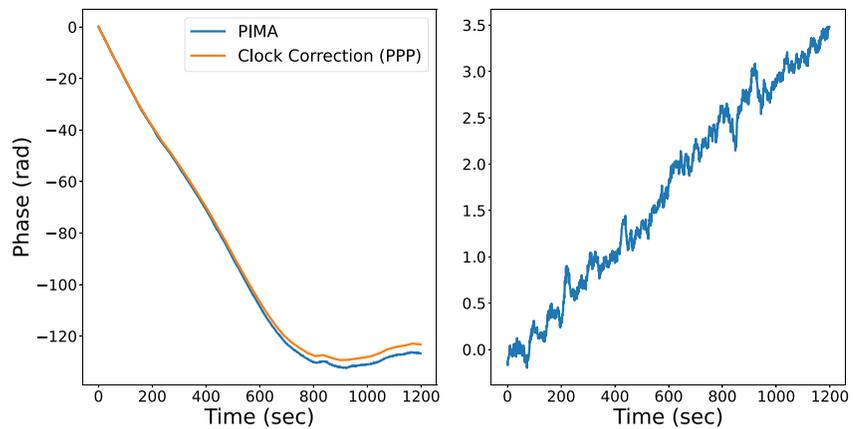

**Figure 11.** (Left) The clock phase drift for FD-GNSS from the empirical fringe fit (blue) and the Precise Point Positioning-derived clock solution (orange) during the 1,200-s 3C 405 scan in the September experiment. (Right) The difference of the two phase drifts, showing a small linear trend indicative of a phase delay rate offset likely due to minor errors in the interferometric model. This remaining trend is corrected during fringe fitting. The RMS phase after a best fit line is removed from the phase differences is 0.1 rad.

### 6.3. System Temperature and SEFD Estimates

To assess the usefulness of the GNSS-radio telescope interferometer, one needs to know its sensitivity. Celestial source detections can be used as a rough check of the SEFD of the modified GNSS antenna using the correlated source flux density $S_f$, the SNR of the detection, and the DPFU and system temperature estimates distributed by NRAO for the FD-VLBA antenna. Equation 8 shows this relationship. The derivation of Equation 8 from Equation 3 is shown in Appendix A.

$$\text{SEFD}_{\text{GNSS}} = 2 \left( \frac{\eta_q S_f}{R_{\text{sn}}} \right)^2 \frac{\text{DPFU}_{\text{VLBA}}}{T_{\text{sys, VLBA}}} \Delta \nu \tau_a \tag{8}$$

In the calculations reported here, we used a quantization efficiency value $\eta_q = 0.8815$ (F. Schwab, 1986), which corresponds to the quantization efficiency of real data optimally digitized at four levels (two bits) and sampled at the Nyquist rate. Although we correlated two-bit sampled data from FD-VLBA against four-bit sampled data from the HRTR, these four-bit data were derived from the original two-bit sampled data with polyphase resampling and therefore cannot restore information lost during the original digitization. The rigorous accounting of the amplitude loss by the re-sampling process is beyond the scope of this work.

Using the assumptions described above, the SEFD of the GNSS antenna can be used to estimate the system temperature with knowledge of the effective antenna area, $A_{\text{GNSS}}$:

$$T_{\text{sys,GNSS}} = \frac{A_{\text{GNSS}}}{2k_B} \text{SEFD}_{\text{GNSS}} \tag{9}$$

The effective antenna area is smaller than the geometric area and accounts for the antenna efficiency. The Topcon CR-G5 GNSS antennas of FD-GNSS and MGO-GNSS have an effective area dependent on the wavelength of the received signal as well as the elevation of the source as shown in Equation 10. The data used in this study to characterize the antenna gain pattern were derived from a full wave simulation of the Topcon CR-G5 antenna gain normalized by setting the boresight gain to the published value for the antenna. The gain data covers 1,156–1,637 MHz in frequency and 0–90 deg in elevation.

**Table 2**
*The Number of Scans and Detections for the Short and Long Baseline by Source Type for All Three Observing Sessions*

| | | Detections | |
|---|---|---|---|
| Source type | Num. scans | FD-VLBA/FD-GNSS | FD-VLBA/MGO-GNSS |
| Natural radio sources | 23 | 22 | 6 |
| GPS | 67 | 67 | 67 |
| Galileo | 3 | 3 | 3 |
| OH Maser | 13 | 0 | 0 |

$$A_{\mathrm{GNSS}}(e, v) = \frac{G(e, v)\lambda^2}{4\pi} \tag{10}$$

Knowledge of the dependence of gain and SEFD on local elevation angle is important for experiment scheduling, as weaker sources may need to be closer to local zenith for a detection of sufficient SNR to permit geodetic analysis.

## 7. Results

The detections by source type for all three observing sessions we conducted in 2022 are shown in Table 2.

### 7.1. Detection of GNSS Sources

Throughout the three observing sessions, we recorded 124 observations of GNSS sources, including both GPS and Galileo satellites, and we have detected all of them. The satellite identifiers and average SNR of the detections are recorded in Table 3 for 5 s coherent accumulations.

We kept coherent accumulations short to avoid unmodeled quadratic phase variations on the longer FD-VLBA/MGO-GNSS baseline. We consistently achieved this goal while providing high SNR detections by using 5 s of coherent accumulation. The detections on the short baseline showed no significant unmodeled phase variation at 40–60 s scan lengths, but are reported in 5 s coherent accumulations for consistency. We confirmed that the scheduled satellite was in the main beam of the VLBA dish by analyzing the satellite signal in autocorrelation spectrum plots.

Figure 12 shows a two-dimensional search window for a GPS satellite detection. The PIMA fringe fitting procedure as given in Equation 4 is a blind search over group delay and phase delay rate. In this paradigm, the search over phase delay rate mimics an acquisition search in Doppler shift for a typical GNSS system, where the main peak and sidelobes of the satellite signal appear as a sinc-squared function on the frequency axis, giving rise to the structure observed on the phase delay rate axis in the figure. The coherent accumulation interval used in this plot is 20 s to show this structure at high SNR for the short baseline.

Figure 13 shows the fringe amplitude and phase for the same GPS satellite detection. The fringe amplitude of a GPS satellite in VLBI processing has a particular structure: the narrow passband amplitude takes the distinctive shape of the summation of GPS L1 broadcast codes. The phase shows low scatter through the passband due to the high signal strength of the GPS source. In this case, a 20 s coherent accumulation resulted in an SNR of 3,778.

The Galileo GNSS system by contrast uses a Binary Offset Carrier (BOC) signal structure, with two main lobes offset from the center carrier frequency with gaps in emission in the passband. The BOC passband of the Galileo signal is clearly shown in the fringe amplitude in Figure 14. The minima in carrier amplitude in the passband lead to regions of relatively unconstrained fringe phase.

### 7.2. Detection of Celestial Sources

The three experiments resulted in 28 successful detections of celestial sources from 46 total observations, with only 1 missed detection on the shorter FD-VLBA–FD-GNSS baseline. In contrast to the GNSS fringes, the

**Table 3**
*Global Navigation Satellite Systems Satellite Detections in the Three Experiments Conducted in 2022 for 5 s Coherent Accumulations*

| SVN/SatID | Date | Num. obs. | Avg. SNR (baseline) | |
|---|---|---|---|---|
| | | | FD-VLBA/FD-GNSS | FD-VLBA/MGO-GNSS |
| 63 | 26 January | 4 | 1,103 | 808 |
| 63 | 21 April | 1 | 598 | 492 |
| 69 | 26 January | 4 | 725 | 613 |
| 74 | 26 January | 1 | 720 | 568 |
| 74 | 21 April | 1 | 2,008 | 567 |
| 50 | 28 September | 1 | 1,461 | 1,417 |
| 48 | 21 April | 4 | 1,309 | 761 |
| 72 | 21 April | 4 | 1,050 | 1,035 |
| 68 | 21 April | 3 | 1,594 | 537 |
| 73 | 26 January | 3 | 828 | 695 |
| 78 | 28 September | 3 | 765 | 559 |
| 58 | 28 September | 1 | 968 | 447 |
| 43 | 21 April | 1 | 320 | 310 |
| 43 | 28 September | 2 | 451 | 394 |
| 77 | 21 April | 1 | 1,684 | 1,488 |
| 56 | 28 September | 2 | 523 | 577 |
| 51 | 28 September | 3 | 695 | 682 |
| 45 | 26 January | 4 | 701 | 348 |
| 45 | 21 April | 2 | 717 | 259 |
| 41 | 26 January | 2 | 849 | 486 |
| 62 | 26 January | 2 | 551 | 342 |
| 62 | 28 September | 2 | 1,211 | 465 |
| 71 | 26 January | 4 | 834 | 798 |
| 66 | 21 April | 2 | 1,230 | 944 |
| 64 | 21 April | 3 | 999 | 543 |
| 52 | 26 January | 4 | 1,540 | 1,424 |
| 70 | 26 January | 3 | 1,535 | 827 |
| GSAT0205 | 28 September | 1 | 913 | 128 |
| GSAT0218 | 28 September | 1 | 402 | 95 |
| GSAT0222 | 28 September | 1 | 515 | 398 |

*Note.* Satellite identifiers "GSATXXXX" are Galileo satellites, and two-digit numbers are GPS Space Vehicle Numbers.

fringes of celestial sources typically appear in the group delay and phase delay rate search as a single large peak surrounded by much smaller sidelobes as shown for the 3C 405/Cygnus A detection in Figure 15.

Celestial sources are broadband, continuum sources in contrast to the narrowband GNSS sources. The fringe phase of a detection of a celestial source should therefore be nearly flat over the full bandwidth given for fringe fitting. The amplitude structure of a successful detection then primarily reflects the shape of the hardware passbands used in receiving the signal. Both of these effects are shown in the fringe phase and amplitude of the 3C 405 detection in Figure 16.

Table 4 shows the sources detected in each experiment along with the SNR of the detection, the local elevation angle of the source at the time of detection, and the integration time and bandwidth used in fringe fitting. The GNSS-VLBA interferometer is susceptible to RFI due to the short baselines we used and the omnidirectionality

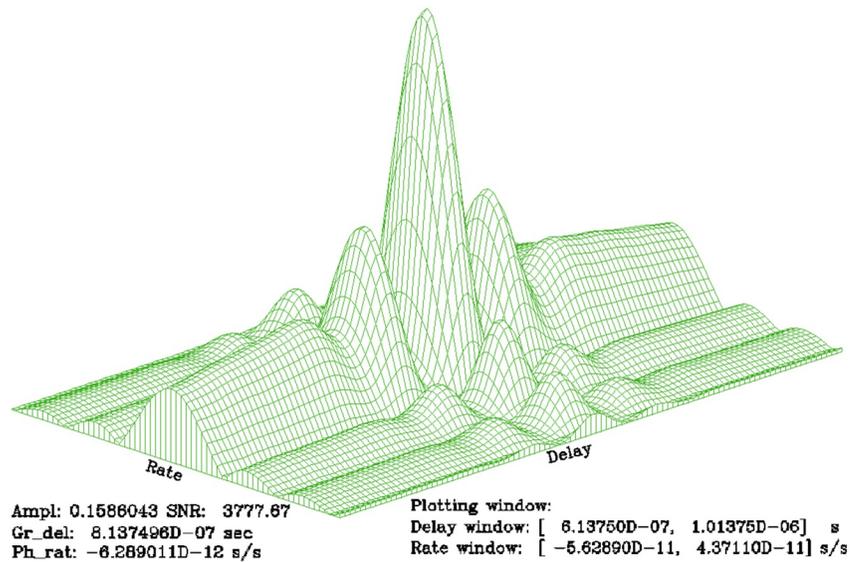

Ampl: 0.1586043  SNR:  3777.87
Gr_del:  8.137496D-07 sec
Ph_rat: -6.289011D-12 s/s

Plotting window:
Delay window: [ 6.13750D-07, 1.01375D-06] s
Rate window: [ -5.62890D-11, 4.37110D-11] s/s

**Figure 12.** The 2-D fringe search for the first observation of Global Positioning System Space Vehicle Number 68 during the 21 April 2022 experiment. This plot shows the results of a 20 s coherent accumulation, which displays the structure of the search window in fine detail. The labels "Ampl" and "SNR" signify the fringe amplitude and the Signal-to-Noise Ratio at the correlation peak. The labels "Gr_del" and "Ph_rat" denote the group delay and phase delay rate, respectively, at which the correlation peak occurs. The fringe search window in group delay and phase delay rate is shown on the bottom right.

of the GNSS antennas. Because of this, in some detections we flagged frequency channels and epochs in which narrowband RFI corrupted the data, reducing the total bandwidth and coherent accumulation time used in fringe fitting. We noticed three detections at abnormally high SNR and with unusual bandpass structure, this amounts to about 10% of our total detections that are problematic. We have not established the nature of this anomaly.

In addition, we carried out a simplified analysis of the observed natural radio sources to characterize the observed flux density and the effect of source structure on the SNR and to support the conclusion that many non-detections of celestial sources on the longer FD-VLBA/MGO-GNSS baseline are due to extended sources being resolved.

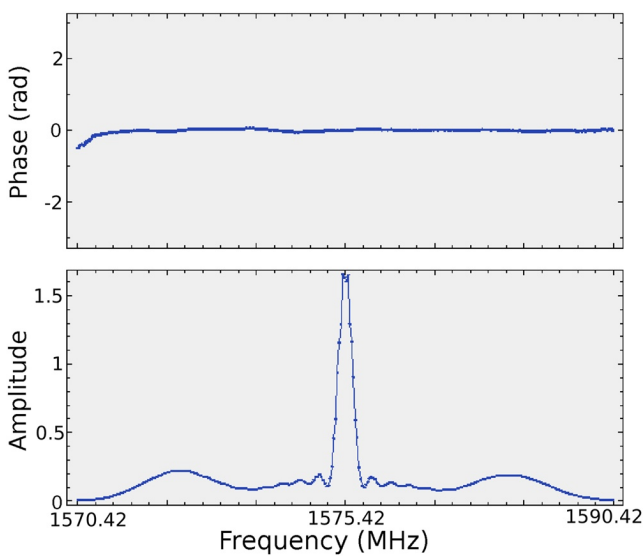

**Figure 13.** The first detection of Global Positioning System Space Vehicle Number 68 on 21 April 2022 with a 20 s coherent accumulation. The top plot shows the fringe phase by frequency, and the bottom plot shows the fringe amplitude by frequency.

Figure 17 demonstrates this analysis, which was inspired by the source structure index described in Charlot (1990) and Tornatore and Charlot (2007).

With a 90 m baseline, the angular resolution is on the order of $\delta\theta \approx \frac{\lambda}{b} = \frac{21\,\text{cm}}{90\,\text{m}} = 8$ arcmin, and the half power beam width of the VLBA antenna is approximately 25 arcmin. For a given source observed in any of the three experiments, we thus query the NRAO VLA Sky Survey (NVSS) catalog for any modeled source components within 10 arcminutes of the observed right ascension and declination. These components are treated as delta functions. The source components are projected to the tangent plane, and through a discrete Fourier transform, we sample the $u$, $v$ plane at 1,000 points equally spaced in time throughout a sidereal day.

The spatial frequencies $u$ and $v$ are obtained from the a priori coordinates of the FD-VLBA radio telescope and the PPP-derived coordinates of the FD-GNSS antenna. Transformations from Earth-Centered, Earth-Fixed (ECEF) coordinates to inertial coordinates are handled by the community-developed Python astronomy package Astropy (Astropy Collaboration et al., 2013, 2018, 2022). The flux density cited in Table 4 for each source detection is the flux density calculated for the spatial coordinates at the time of the observation.

Over one sidereal day, a single baseline traces an ellipse in the $u$-$v$ plane. We sample this ellipse at 1,000 points equally spaced in time, and from these 1,000 points we calculate the average flux density and the standard deviation of the samples in janskys. Comparing the average and standard deviation calculated from all possible spatial frequencies for the given baseline to the flux density at the observation time in Table 4 gives a rough indication of how significant the effect of source structure is for each source. The larger the difference between the average flux density and the given flux density and the closer the magnitude of the standard deviation of the flux density along the ellipse to the average, the more pronounced the effect of the source's extended structure.

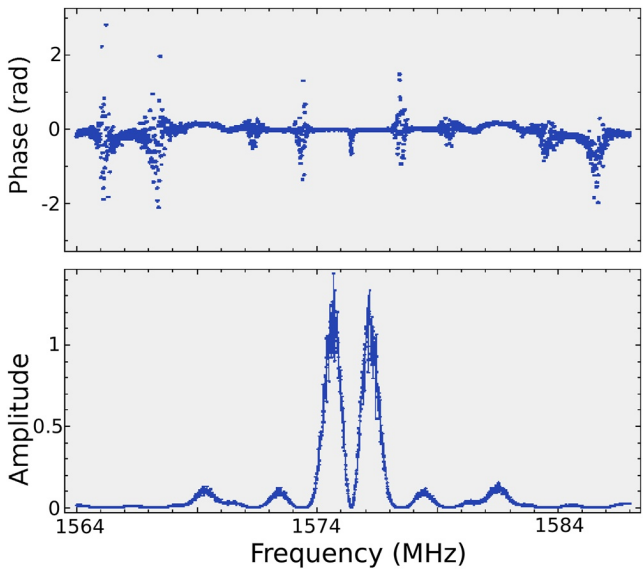

**Figure 14.** The fringe amplitude and phase by frequency for the observation of Galileo satellite GSAT0205 in the 28 September 2022 experiment. The coherent accumulation time is 20 s. The Binary Offset Carrier signal structure used by Galileo means that there are frequency regions in the bandpass with no signal, causing increased phase noise.

### 7.3. Characterization of Sensitivity

The detections of the source 3C 295, a stable calibrator source observed in each experiment, are of particular interest in characterizing the sensitivity of the GNSS-VLBA interferometer. The gain of the Topcon CR-G5 antenna at the center frequency of 1,440 MHz is shown as a function of local elevation angle in Figure 18 with the 3C 295 observations labeled.

Table 5 provides interpolated gain values and corresponding system temperatures as calculated from Equation 9 for the 3C 295 observations, where the assumed correlated flux density is that given in the NVSS catalog, 22.7 Jy. For the January observing session, we used only the first HRTR band in calculating the sensitivity because the second band was placed at the border of the two 128 MHz VLBA bands, causing an uncharacteristic drop in fringe amplitude that corrupts the sensitivity measurement. Thus, the measured SNR is 31 with a bandwidth of 40 MHz rather than the 41 SNR and 80 MHz reported in Table 4.

The calculated system temperature for the 21 April 2022 observation of 3C 295 is of lower quality because the gain pattern is not as well determined for low elevations, and ground effects may influence the apparent received emission strength. Using the calculated system temperatures for the January and September observing sessions, the average system temperature is about 330 K. Assuming a system temperature for the GNSS receiving system of 330 K, we can make predictions for the SEFD of the GNSS-VLBA interferometer and thus the number of sources observable under different conditions. Equation 11 shows the SEFD of the combined interferometer, which can

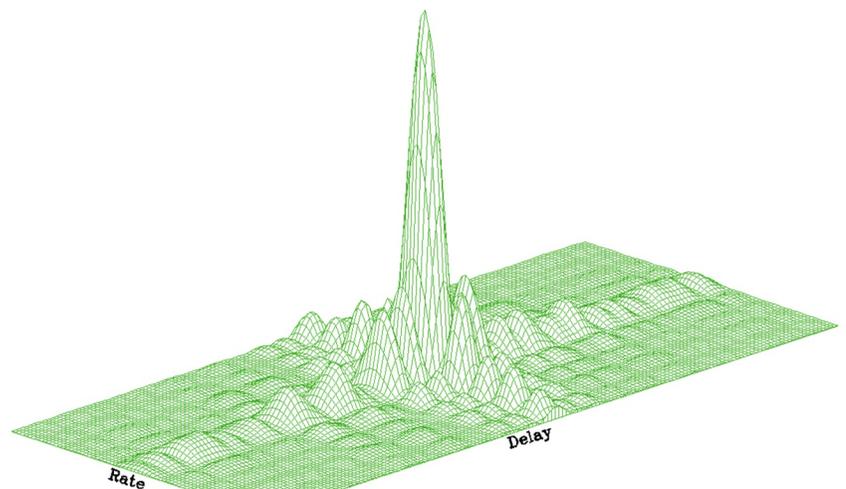

```
Ampl: 0.0044516 SNR: 1717.29   Plotting window:
Gr_del: 9.085173D-07 sec        Delay window: [ 8.33517D-07, 9.83517D-07]  s
Ph_rat: 3.553554D-13 s/s        Rate window:  [-1.14464D-12, 1.85536D-12] s/s
```

**Figure 15.** The 2-D fringe search for the observation of 3C 405 during the 21 April 2022 experiment. The source is detected with very high Signal-to-Noise Ratio and shows a clear central peak with comparatively small sidelobes.

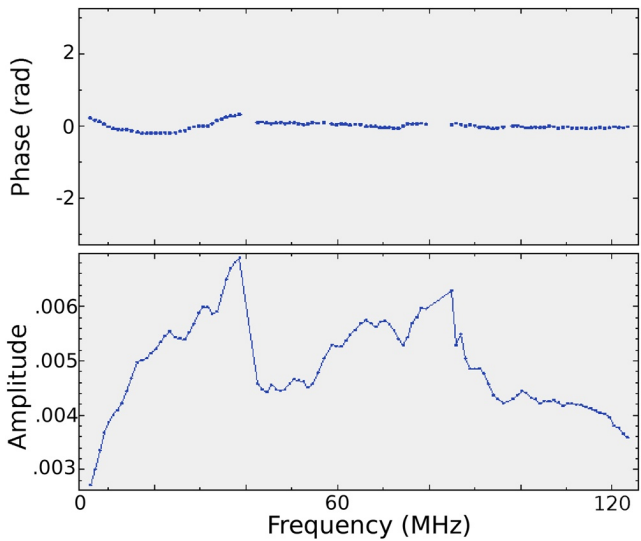

**Figure 16.** The fringe amplitude and phase by frequency for the observation of 3C 405 in the 21 April 2022 experiment. The fringe phase is nearly flat, and the fringe amplitude is determined mainly by roll-off due to hardware and software filters. The frequencies shown are not contiguous but represent the approximately 36 MHz-wide bands used in the correlation.

be obtained as a geometric mean of the system temperatures and effective areas of the elements.

$$\text{SEFD} = \frac{2k_B\sqrt{T_{\text{sys,GNSS}}T_{\text{sys,VLBA}}}}{\sqrt{A_{\text{GNSS}}A_{\text{VLBA}}}} = \sqrt{\text{SEFD}_{\text{VLBA}}\text{SEFD}_{\text{GNSS}}} \quad (11)$$

Figure 19 shows the SEFD for the elements of the GNSS-VLBA interferometer. The SEFD of the VLBA radio telescope is constant with respect to elevation and is distributed by NRAO. The GNSS and combined interferometer SEFD are computed with the elevation-dependent gain pattern of the GNSS antenna.

For a GNSS system temperature of 330 K, a bandwidth of 200 MHz, and 5 min integration time at 45° elevation, the flux density required for a detection of 10 SNR according to Equations 8 and 9 is 6.4 Jy. These numbers represent a reasonable midpoint—200 MHz bandwidth is between 5 and 6 of the total 9 configurable HRTR bands assumed to be tuned to useable, relatively radio quiet frequencies. 5 minutes of integration time, while longer than a typical scan in geodetic experiments, represents a reasonable scan length for producing high SNR detections on many sources in a full experiment. A typical elevation angle for an observed source is 45°, although for estimation of important geodetic parameters, a variety of source declinations and therefore elevations is required.

Considering the point source NVSS catalog and combining the flux density of sources within 10 arcminutes of each other as would be observed on a short baseline with coarse angular resolution, there are more than 150 sources brighter than this 6.4 Jy limiting flux density, suggesting that geodetic experiments are feasible for the GNSS-VLBA interferometer with celestial sources as the primary target.

### 7.4. Search for OH Masers

We processed the data for 13 total scans of nine unique sources including one molecular cloud and eight stellar OH masers in the 28 September 2022 experiment. We found a strong indication of the presence of one or more of the hydroxide emission lines at 1,612, 1,665, and 1,667 MHz in the FD-VLBA autocorrelation for eight of these 13 scans, corresponding to five individual masers (VY CMa, W3(OH), OH138.0 + 07.2, BW Cam, IRAS 07331 + 0021). However, we have not detected these emission lines in any cross-correlation spectra. This suggests inadequate sensitivity for a detection of the selected OH masers, but because we cannot rule out other confounding factors with a single recording session, the data is insufficient to conclusively determine if this is true of OH masers on VLBI/GNSS baselines more generally.

## 8. Concluding Remarks

We have developed a technology that allows for a GNSS antenna and receiver to be used as an element of a radio interferometer. We used common VLBI processing tools to analyze our data, and we successfully produced a strong interferometric response on both natural radio sources and GNSS satellites. We ran three experiments, each 3–4.5 hr long, between two HRTRs and a 25 m FD-VLBA antenna with baselines of about 100 and 9,000 m length. As expected, we have easily detected signals from GPS and Galileo navigation satellites. We found that it is sufficient to record for 5 s to reach an SNR of about 1,000. We also observed 12 natural radio sources—three Galactic supernova remnants, and nine AGNs. We have detected all of them with an SNR in a range of 20–1,720. We attempted to observe Galactic OH masers, but were unable to detect them. Although these sources are not typically used in geodetic VLBI, we attempted to observe them as a method of unlocking an additional source type for low-sensitivity radio interferometry. Our results suggest that we may not have the sensitivity to detect OH masers, but our data were not fully conclusive as to whether this is the case generally.

**Table 4**
*Celestial Source Detections in the Three Conducted Experiments*

| Source | GNSS station | Date | Elev. (deg) | Int. time (sec) | Bandwidth (MHz) | SNR | Flux dens. (Jy) | Ellipse analysis Avg. flux dens. (Jy) | $\sigma$ (Jy) |
|---|---|---|---|---|---|---|---|---|---|
| 3C 84* | FD | 21 April | 28 | 650 | 110 | 368 | 22.9 | 22.8 | 0.0 |
| 3C 84 | FD | 28 September | 37 | 600 | 80 | 36 | 22.8 | 22.8 | 0.0 |
| 3C 123 | FD | 28 September | 55 | 602 | 120 | 84 | 49.8 | 49.7 | 0.1 |
| 3C 144 | FD | 28 September | 63 | 600 | 120 | 630 | 750.1 | 719.1 | 95.8 |
| 3C 218 | FD | 28 September | 44 | 600 | 80 | 63 | 40.9 | 40.9 | 0.1 |
| 3C 273* | FD | 28 September | 24 | 600 | 120 | 768 | 55.2 | 55.1 | 0.2 |
| 3C 274 | FD | 26 January | 55 | 390 | 80 | 110 | 135.0 | 138.1 | 3.9 |
| 3C 274 | FD | 26 January | 22 | 600 | 80 | 65 | 136.2 | 138.1 | 3.9 |
| 3C 274* | FD | 28 September | 31 | 330 | 120 | 386 | 132.6 | 138.1 | 3.7 |
| 3C 295 | FD | 26 January | 67 | 600 | 80 | 41 | 22.7 | 22.7 | 0.0 |
| 3C 295 | FD | 21 April | 10 | 600 | 110 | 21 | 22.7 | 22.7 | 0.0 |
| 3C 295 | FD | 28 September | 21 | 600 | 120 | 23 | 22.7 | 22.7 | 0.0 |
| 3C 348 | FD | 26 January | 64 | 600 | 60 | 56 | 38.7 | 40.9 | 4.3 |
| 3C 348 | FD | 21 April | 17 | 600 | 110 | 43 | 46.6 | 40.9 | 4.3 |
| 3C 353 | FD | 26 January | 45 | 440 | 80 | 32 | 34.9 | 43.4 | 8.0 |
| 3C 353 | FD | 26 January | 58 | 600 | 60 | 38 | 31.7 | 43.4 | 8.0 |
| 3C 353 | FD | 21 April | 22 | 600 | 110 | 37 | 52.8 | 43.4 | 8.0 |
| 3C 398 | FD | 26 January | 50 | 600 | 80 | 29 | 36.2 | 36.6 | 3.8 |
| 3C 398 | FD | 21 April | 58 | 602 | 110 | 119 | 37.5 | 36.6 | 3.8 |
| 3C 400 | FD | 26 January | 48 | 602 | 80 | 43 | 46.7 | 47.6 | 8.9 |
| 3C 400 | FD | 21 April | 59 | 600 | 110 | 84 | 46.9 | 47.6 | 8.9 |
| 3C 405 | FD | 21 April | 73 | 1,200 | 110 | 1,719 | 1,493.8 | 1,458.2 | 112.1 |
| 3C 84 | MGO | 21 April | 28 | 650 | 110 | 12 | | | |
| 3C 84 | MGO | 28 September | 37 | 600 | 120 | 18 | | | |
| 3C 273 | MGO | 26 January | 50 | 600 | 80 | 47 | | | |
| 3C 273 | MGO | 28 September | 24 | 600 | 120 | 25 | | | |
| 3C 295 | MGO | 26 January | 68 | 600 | 80 | 26 | | | |
| 3C 405 | MGO | 21 April | 73 | 1,200 | 110 | 93 | | | |

*Note.* Source detections marked with * indicate that the detection showed unusual bandpass features and abnormally high fringe amplitude. In the ellipse analysis, we included source components within 10 arcminutes, and the average flux density is computed along the ellipse in spatial frequency formed over a sidereal day for the short baseline. $\sigma$ is the standard deviation of flux density along the ellipse in spatial frequency. A standard deviation $\sigma = 0.0$ Jy indicates that there is only a single bright source component in the included field of view. We do not include the ellipse analysis for the longer baseline MGO-GNSS detections because the point flux densities in the catalog are not representative for a 9,000 m baseline length.

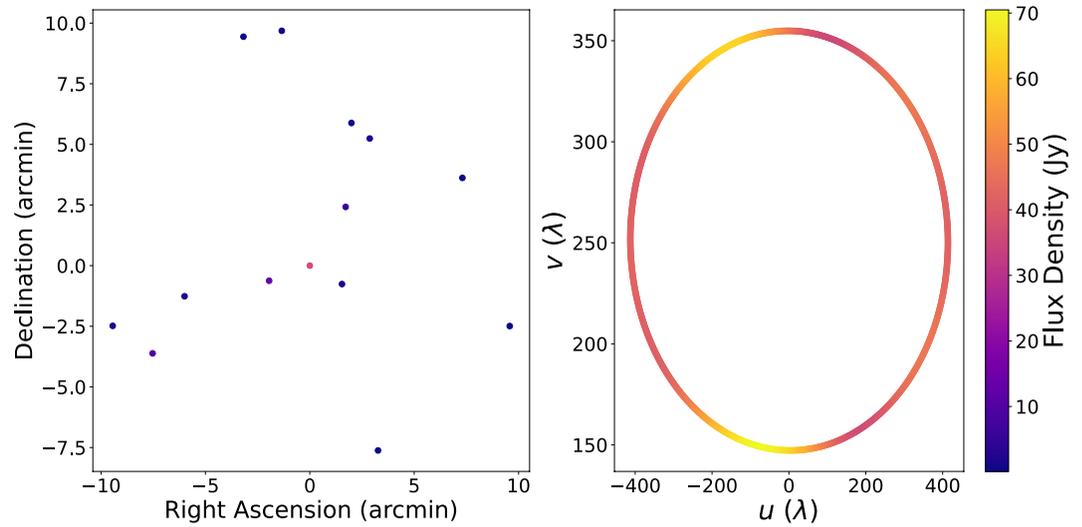

**Figure 17.** The National Radio Astronomy Observatories Very Large Array Sky Survey distribution of point sources on the sky for the extended source 3C 400 (left) and the corresponding ellipse in spatial frequency drawn over a sidereal day for the short baseline (right). The flux density in janskys is given on the color axis for both plots.

Using the NVSS catalog, we find that more than 150 sources can be detected at short baselines with 5 min of integration time at 1.5 GHz. This estimate is based on a zenith direction SEFD of the GNSS antenna of about 60 MJy, which leads to a limiting flux density of 6.4 Jy for a radio source at an elevation angle of 45°. The GNSS/FD-VLBA radio interferometer has a baseline sensitivity of ~130 kJy. This is a factor of 50 worse than between two geodetic 12 m VLBI Global Observing System (VGOS) antennas.

A major application of this system is the measurement of local ties between GNSS and VLBA antennas. The accuracy of these measurements based on phase delays may reach a level below 1 mm if modeling errors are constrained to the levels seen in prior radio telescope to radio telescope experiments. These measurements provide the positions of microwave reference points with respect to each other directly. Another potential application is that the clock determination technique explored here might enable time transfer to VLBI stations through GNSS-radio telescope observations.

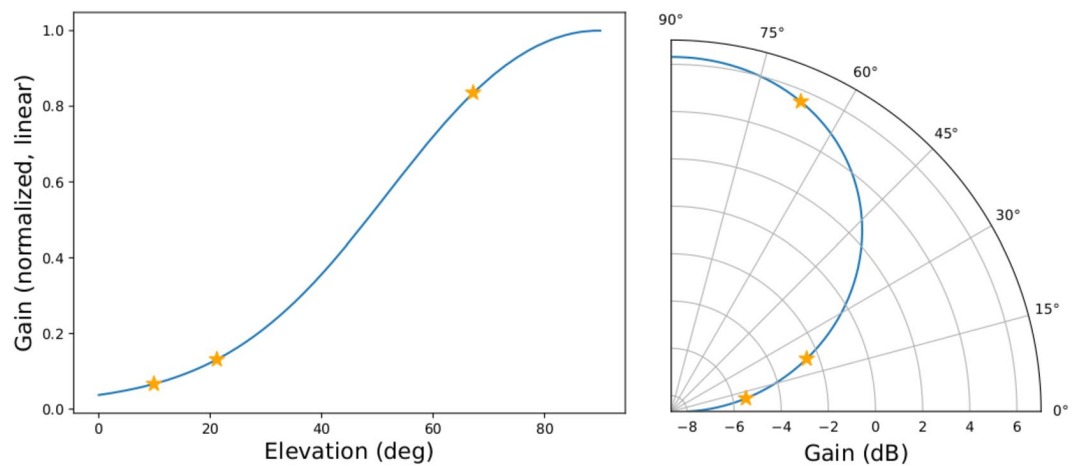

**Figure 18.** The gain pattern of the Topcon CR-G5 antenna with 3C 295 observations labeled. (Left) The gain pattern in normalized linear units as a function of elevation. (Right) The radial gain pattern in dB. Note that both plots contain the same information.

**Table 5**
*Computed System Equivalent Flux Density and System Temperatures for the Global Navigation Satellite Systems Antenna From Detections of the Compact Source 3C 295 on Baseline FD-VLBA–FD-GNSS in the Three Data Collections at Fort Davis*

| Experiment | Elevation (deg) | Gain (dB) | SNR | SEFD (MJy) | System temp. (K) |
|---|---|---|---|---|---|
| 26 January 2022 | 67 | 5.5 | 31 | 60.5 | 359 |
| 21 April 2022 | 10 | −5.5 | 21 | 29.6 | 160 |
| 28 September 2022 | 21 | −2.4 | 23 | 55.0 | 296 |

Future work will focus on the development of this technique for geodetic applications including the measurement, characterization, and optimization of local tie vectors. To do this, we plan to develop an automated pipeline for data analysis that will produce geodetic grade group and phase delays from longer experiments with monumented GNSS antennas and stable clock references. Processing these geodetic observables from GNSS satellites will require the implementation of a rigorous near-field path delay model such as the one described in Jaron and Nothnagel (2019). We will also implement an advanced procedure for flagging visibilities affected by RFI. Finally, in order to contribute to the determination of terrestrial references frames, we will need to determine tie vectors between VLBI antennas and IGS reference stations that operationally contribute to multi-technique combination reference frames.

It has long been known that placing a GNSS receiver in the vicinity of a VLBI antenna is beneficial because the GNSS receiver senses the same atmosphere and can be used for calibrating VLBI data (Ros et al., 2000). With the technology that we have demonstrated, we are in a position to further exploit the synergism of these two techniques. With the vector ties known with a sub-millimeter accuracy, we can consider a GNSS receiving system and radio telescope a single instrument with enhanced capabilities. We can analyze data separately and produce the differences between VLBI position time series and GNSS time series referred to the VLBI reference point for investigation of the systematic errors of these two techniques.

The system that we have developed costs far less than traditional VLBI systems, is transportable, and has a great potential for outreach activities. Early digitization in the HRTR allows us to shift the focus of our work to processing data records and substantially reduces the barrier to entry for the VLBI community.

Our GNSS antenna pre-amplifier and filter modifications are sufficient to increase the passband of the antenna, allowing access to a greater segment of the L band. We are currently developing a permanent, integrated solution for the passband modifications implemented here with connectorized components. Once this is complete, we plan to obtain calibration measurements of the modified antenna. We can use the data collected by the modified GNSS antenna and HRTR after limited pre-processing as a data stream from a radio telescope, allowing us to use existing VLBI data analysis software and tools.

By leveraging simultaneous collection of traditional GPS observables to compute a clock solution, we have also demonstrated the ability to perform long coherent accumulations with a rubidium frequency standard. In our longest scan of 1,200 s, we found no measurable degradation of the SNR with respect to its expected growth as the square root of integration time. Therefore, we can only provide the lower limit of decorrelation caused by clock instability after the correction is applied. We do not yet know the upper limit of the integration time that our technique can achieve, but these results are very promising and warrant further investigation.

In addition, HRTR systems are currently planned for deployment at all VLBA stations in support of the National Geodetic Survey Continuously Operating Reference System network (York, 2021). This raises the possibility that short baseline observations with HRTR systems in a GNSS-radio telescope configuration may be possible on a wider scale, and these HRTR GNSS systems should be considered for use with the techniques proposed once deployed. Prospective modifications of the existing VGOS signal chain to add the capability to record L band signal from navigation satellites following the ideas of Kodet et al. (2014) present another possibility for GNSS-radio telescope observations. We envisage that all radio telescopes in the future will have a collocated GNSS receiver with the capability to record baseband data.

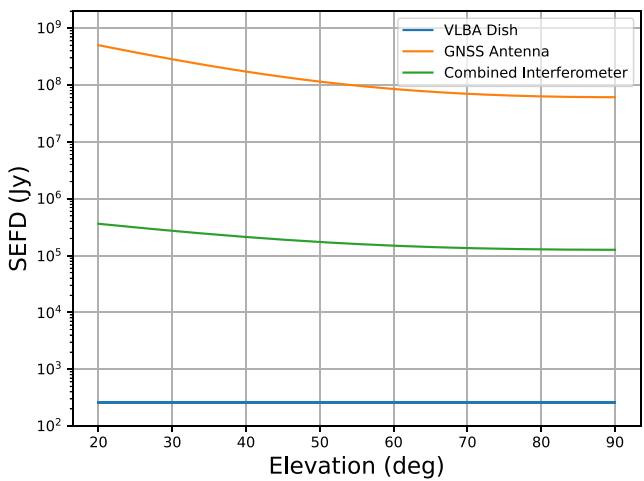

**Figure 19.** The System Equivalent Flux Density of the Very Long Baseline Array (VLBA) radio telescope (blue), Global Navigation Satellite Systems (GNSS) antenna (orange), and GNSS-VLBA interferometer (green) versus elevation, assuming $T_{sys,GNSS} = 330$ K.

## Appendix A: Derivation of SEFD$_{\text{GNSS}}$ Expression

Starting from Equation 3, the antenna temperature can be expressed in terms of the source flux density and System Equivalent Flux Density (SEFD):

$$T_{\text{ant}} = T_{\text{sys}} \frac{S_f}{\text{SEFD}} \tag{A1}$$

Substitute this expression into Equation 3 for the Very Long Baseline Array (VLBA) radio telescope and Global Navigation Satellite Systems (GNSS) antenna:

$$R_{\text{sn}} = \eta_q \sqrt{\frac{S_f^2}{\text{SEFD}_{\text{GNSS}} \text{SEFD}_{\text{VLBA}}} 2 \Delta \nu \tau_a} \tag{A2}$$

SEFD is a function of system temperature and Degrees per Flux Unit (DPFU) and is elevation independent for the L band VLBA receiving system:

$$\text{SEFD}_{\text{VLBA}} = \frac{T_{\text{sys,VLBA}}}{\text{DPFU}} \tag{A3}$$

Substituting for SEFD$_{\text{VLBA}}$, we get,

$$R_{\text{sn}} = \eta_q \sqrt{\frac{S_f \text{DPFU}}{\text{SEFD}_{\text{GNSS}} T_{\text{sys, VLBA}}} 2 \Delta \nu \tau_a} \tag{A4}$$

Finally, solving for SEFD$_{\text{GNSS}}$, we obtain Equation 8. Further, from Equation 9 the system temperature $T_{\text{sys,GNSS}}$ assuming no data loss is given by,

$$T_{\text{sys,GNSS}} = \left( \frac{\eta_q S_f}{R_{\text{sn}}} \right)^2 \frac{\text{DPFU}_{\text{VLBA}}}{k_B T_{\text{sys,VLBA}}} A_{\text{GNSS}} \Delta \nu \tau_a \tag{A5}$$

## Acronyms

| | |
|---|---|
| AGN | Active Galactic Nuclei |
| BOC | Binary Offset Carrier |
| CIC | Cascaded Integrator-Comb |
| CORS | Continuously Operating Reference System |
| DPFU | Degrees per Flux Unit |
| DiFX | Distributed FX |
| ECEF | Earth-Centered, Earth-Fixed |
| FD | Fort Davis, Texas |
| FIR | Finite Impulse Response |
| FITS | Flexible Image Transport System |
| FPGA | Field Programmable Gate Array |
| GNSS | Global Navigation Satellite Systems |
| GPS | Global Positioning System |
| GSPS | Gigasamples per Second |
| HDF5 | Hierarchical Data Format, version 5 |
| HRTR | High Rate Tracking Receiver |
| IGS | International Global Navigation Satellite Systems (GNSS) Service |
| MGO | McDonald Geodetic Observatory |
| MSPS | Megasamples per Second |
| NGS | National Geodetic Survey |
| NORAD | North American Aerospace Defense Command |

| NRAO | National Radio Astronomy Observatories |
| NVSS | National Radio Astronomy Observatories (NRAO) Very Large Array (VLA) Sky Survey |
| PPP | Precise Point Positioning |
| RFI | Radio-Frequency Interference |
| SEFD | System Equivalent Flux Density |
| SNR | Signal-to-Noise Ratio |
| SVN | Space Vehicle Number |
| TLE | Two-Line Element |
| VDIF | Very Long Baseline Interferometry (VLBI) Data Interchange Format |
| VGOS | VLBI Global Observing System |
| VLA | Very Large Array |
| VLBA | Very Long Baseline Array |
| VLBI | Very Long Baseline Interferometry |

## Data Availability Statement

All FITS data produced from the described experiments as well as custom code used in producing the plots in this manuscript are available at the Texas Data Repository (Skeens et al., 2023). GPS final orbit products including ephemerides and clock solutions were downloaded from the Crustal Dynamics Data Information System (NASA Crustal Dynamics Data Information System, 1992; Noll, 2010). Figures showing fringe fitting data were produced with PGPLOT (Pearson, 2011). All other figures in this manuscript were produced with Matplotlib version 3.5.3 (Caswell et al., 2022; Hunter, 2007). Release 9 January 2023 of the fringe fitting software PIMA is available from http://astrogeo.org/pima (Petrov, 2023; Petrov et al., 2011) and is continuously developed. Version 2.6.2 of the software correlator DiFX is available at https://svn.atnf.csiro.au/difx/master_tags/DiFX-2.6.2 (A. Deller, 2022; A. T. Deller et al., 2011).

**Acknowledgments**
The authors thank Walter Brisken at NRAO for help in coordinating these observations, Sharyl Byram for sponsoring USNO observing time, and the Fort Davis NRAO technicians Julian Wheat and Juan De Guia who provided operational support. We would also like to thank Eusebio "Chevo" Terrazas and Renny Spencer at McDonald Observatory for the time and assistance they provided during data collections. We are grateful to Mark Claussen of NRAO for the suggestion to investigate hydroxide masers as potential sources for the interferometer as well as advice in maser observing and to Dave Rainwater of ARL:UT, who reviewed the manuscript and provided many helpful comments. We gratefully acknowledge the funding provided for this work under NASA Grant 80NSSC20K1732 and NASA IIP Grant NNX17AD29G. ARL:UT Research and Development supported work not covered under the NASA grants. The NRAO is a facility of the National Science Foundation operated under cooperative agreement by Associated Universities, Inc. The authors acknowledge use of the VLBA under the USNO's time allocation. This work made use of the Swinburne University of Technology software correlator, developed as part of the Australian Major National Research Facilities Programme and operated under license.

## References

Astropy Collaboration, Price-Whelan, A. M., Lim, P. L., Earl, N., Starkman, N., Bradley, L., et al. (2022). The Astropy project: Sustaining and growing a community-oriented open-source project and the latest major release (v5.0) of the core package. *The Astrophysical Journal, 935*(2), 167. https://doi.org/10.3847/1538-4357/ac7c74

Astropy Collaboration, Price-Whelan, A. M., Sipőcz, B. M., Günther, H. M., Lim, P. L., Crawford, S. M., et al. (2018). The Astropy project: Building an open-science project and status of the v2.0 core package. *The Astronomical Journal, 156*(3), 123. https://doi.org/10.3847/1538-3881/aabc4f

Astropy Collaboration, Robitaille, T. P., Tollerud, E. J., Greenfield, P., Droettboom, M., Bray, E., et al. (2013). Astropy: A community Python package for astronomy. *Astronomy & Astrophysics, 558*, A33. https://doi.org/10.1051/0004-6361/201322068

Broten, N. W., Locke, J. L., Legg, T. H., McLeish, C. W., Richards, R. S., Chisholm, R. M., et al. (1967). Observations of quasars using interferometer baselines up to 3,074 km. *Nature, 215*(5096), 38. https://doi.org/10.1038/215038a0

Brown, G. W., Carr, T. D., & Block, W. F. (1968). Long-baseline interferometry of S-bursts from Jupiter. *Astrophysical Letters, 1*, 89.

Carter, W. E., Rogers, A. E. E., Counselman, C. C., & Shapiro, I. I. (1980). Comparison of geodetic and radio interferometric measurements of the Haystack-Westford base line vector. *Journal of Geophysical Research, 85*(B5), 2685–2687. https://doi.org/10.1029/JB085iB05p02685

Caswell, T. A., Lee, A., Droettboom, M., de Andrade, E. S., Hoffmann, T., Klymak, J., et al. (2022). matplotlib/matplotlib: REL: V3.5.3 [Software]. Zenodo. https://doi.org/10.5281/zenodo.6982547

Charlot, P. (1990). Radio-source structure in astrometric and geodetic very long baseline interferometry. *The Astronomical Journal, 99*, 1309. https://doi.org/10.1086/115419

Cohen, R. J. (1989). Compact maser sources. *Reports on Progress in Physics, 52*(8), 881–943. https://doi.org/10.1088/0034-4885/52/8/001

Deller, A. T. (2022). DiFX 2.6.2 [Software]. ATNF. Retrieved from https://svn.atnf.csiro.au/difx/master_tags/DiFX-2.6.2/

Deller, A. T., Brisken, W. F., Phillips, C. J., Morgan, J., Alef, W., Cappallo, R., et al. (2011). DiFX-2: A more flexible, efficient, robust, and powerful software correlator. *Publications of the Astronomical Society of the Pacific, 123*(901), 275–287. https://doi.org/10.1086/658907

Deller, A. T., Tingay, S. J., Bailes, M., & West, C. (2007). DiFX: A software correlator for very long baseline interferometry using multiprocessor computing environments. *Publications of the Astronomical Society of the Pacific, 119*(853), 318–336. https://doi.org/10.1086/513572

Dieck, C., Johnson, M. C., & MacMillan, D. S. (2023). The importance of co-located VLBI Intensive stations and GNSS receivers. *Journal of Geodesy, 97*(3), 21. https://doi.org/10.1007/s00190-022-01690-1

Engels, D., & Bunzel, F. (2015). A database of circumstellar oh masers. *Astronomy & Astrophysics, 582*, A68. https://doi.org/10.1051/0004-6361/201322589

Greisen, E. (2022). *The FITS interferometer data Interchange convention—Revised (technical report) (AIPS Memo 114)*. National Radio Astronomy Observatory. Retrieved from http://www.aips.nrao.edu/TEXT/PUBL/AIPSMEM114.PDF

Hase, H., & Petrov, L. (1999). The first campaign of observations with the VLBI-module of TIGO. In W. Schlüter & H. Hase (Eds.), *Proceedings of the 13th working meeting on European VLBI for geodesy and astrometry* (pp. 19–24). BKG.

Hellerschmied, A., Böhm, J., Kwak, Y., McCallum, J., & Plank, L. (2016). VLBI observations of GNSS signals on the baseline Hobart-Ceduna. In D. Behrend, K. Baver, & K. Armstrong (Eds.), *IVS 2016 general meeting proceedings* (pp. 373–376).

Herring, T. (1992). Submillimeter horizontal position determination using very long baseline interferometry. *Journal of Geophysical Research, 97*(B2), 1981–1990. https://doi.org/10.1029/91JB02649

Hunter, J. D. (2007). Matplotlib: A 2D graphics environment. *Computing in Science & Engineering, 9*(3), 90–95. https://doi.org/10.1109/MCSE.2007.55

Jaron, F., & Nothnagel, A. (2019). Modeling the VLBI delay for earth satellites. *Journal of Geodesy, 93*(7), 953–961. https://doi.org/10.1007/s00190-018-1217-0

Kodet, J., Schreiber, K. U., Plötz, C., Neidhardt, A., Kronschnabl, G., Haas, R., et al. (2014). Co-locations of space geodetic techniques on ground and in space. In *International VLBI service for geodesy and astrometry 2014 general meeting proceedings: VGOS: The new VLBI network* (pp. 446–450).

Matsumoto, S., Ueshiba, H., Nakakuki, T., Takagi, Y., Hayashi, K., Yutsudo, T., et al. (2022). An effective approach for accurate estimation of VLBI-GNSS local-tie vectors. *Earth Planets and Space, 74*(147), 147. https://doi.org/10.1186/s40623-022-01703-5

Matveenko, L. I., Kardashev, N.-S., & Sholomitskii, G.-B. (1965). Large base-line radio interferometers. *SvRP, 461*(4), 461–463. https://doi.org/10.1007/bf01038318

Napier, E. A., & Peter, J. (1994). The Very Long Baseline Array. *Proceedings of the IEEE, 82*(5), 658–672. https://doi.org/10.1109/5.284733

NASA Crustal Dynamics Data Information System. (1992). GNSS final combined orbit solution product [Dataset]. NASA. https://doi.org/10.5067/GNSS/GNSS_IGSORB_001

Niell, A., Barrett, J., Cappallo, R., Corey, B., Elosegui, P., Mondal, D., et al. (2021). VLBI measurement of the vector baseline between geodetic antennas at Kokee Park Geophysical Observatory, Hawaii. *Journal of Geodesy, 95*(6), 65. https://doi.org/10.1007/s00190-021-01505-9

Ning, T., Haas, R., & Elgered, G. (2015). Determination of the local tie vector between the VLBI and GNSS reference points at Onsala using GPS measurements. *Journal of Geodesy, 89*(7), 711–723. https://doi.org/10.1007/s00190-015-0809-1

Noll, C. E. (2010). The crustal dynamics data information system: A resource to support scientific analysis using space geodesy. *Advances in Space Research, 45*(12), 1421–1440. https://doi.org/10.1016/j.asr.2010.01.018

Olson, C. G., & Tolman, B. W. (2018). Robust GNSS differential processing for all baselines. In *Proceedings of the 31st international technical meeting of the satellite division of the institute of navigation (ION GNSS + 2018), Miami, FL* (pp. 3927–3944). https://doi.org/10.33012/2018.16041

Pearson, T. (2011). PGPLOT: Device-Independent graphics package for simple scientific graphs [Software]. Astrophysics Source Code Library. Retrieved from https://ascl.net/1108.003

Petrov, L. (2021). The wide-field VLBA calibrator survey: WFCS. *Astronomical Journal, 161*(1), 14. https://doi.org/10.3847/1538-3881/abc4e1

Petrov, L. (2023). PIMA 2023-01-09. [Software]. Astrogeo Center. Retrieved from http://astrogeo.org/pima

Petrov, L., Kovalev, Y., Fomalont, E., & Gordon, D. (2011). The very long baseline Array galactic plane survey-VGaPS. *The Astronomical Journal, 142*(2), 35. https://doi.org/10.1088/0004-6256/142/2/35

Petrov, L., York, J., Skeens, J., Ji-Cathriner, R., Munton, D., & Herrity, K. (2023). Precise VLBI/GNSS ties with micro-VLBI. (arXiv:2302.05555).

Plank, L., Hellerschmied, A., McCallum, J., Böhm, J., & Lovell, J. (2017). VLBI observations of GNSS-satellites: From scheduling to analysis. *Journal of Geodynamics, 91*(7), 867–880. https://doi.org/10.1007/s00190-016-0992-8

Ray, J., & Altamimi, Z. (2005). Evaluation of co-location ties relating the VLBI and GPS reference frames. *Journal of Geodesy, 79*(4–5), 189–195. https://doi.org/10.1007/s00190-005-0456-z

Rogers, A. E. E., Knight, C. A., Hinteregger, H. F., Whitney, A. R., Counselman, C. C., Shapiro, I. I., et al. (1978). Geodesy by radio interferometry: Determination of a 1.24-km base line vector with ~5-mm repeatability. *Journal of Geophysical Research, 83*(B1), 325–334. https://doi.org/10.1029/JB083iB01p00325

Ros, E., Marcaide, J. M., Guirado, J. C., Sardón, E., & Shapiro, I. I. (2000). A GPS-based method to model the plasma effects in VLBI observations. *Astronomy & Astrophysics, 356*, 357–362.

Schwab, F. (1986). *Two-bit correlators: Miscellaneous results (technical report) (VLBA correlator Memo 75)*. National Radio Astronomy Observatory. Retrieved from https://library.nrao.edu/public/memos/vlba/corr/VLBAC_75.pdf

Skeens, J., York, J., Petrov, L., Munton, D., Herrity, K., Ji-Cathriner, R., et al. (2023). First observations with a GNSS antenna to radio telescope interferometer [Dataset]. Texas Data Repository.

Thompson, J. M., Moran, A. R., & Swenson, G. W., Jr. (2001). *Interferometry and synthesis in radio astronomy* (2nd ed.). John Wiley and Sons.

Tornatore, V., Haas, R., Casey, S., Duev, D., Pogrebenko, S., & Calvés, G. M. (2014). Direct VLBI observations of global navigation satellite systems. In C. Rizos & P. Willis (Eds.), *Earth on the edge: Science for a sustainable planet* (Vol. 139, pp. 247–252). Springer-Verlag.

Tornatore, V., Haas, R., & Nilsson, T. (2007). The impact of radio source structure on European geodetic VLBI measurements. *Journal of Geodesy, 81*(6), 469–478. https://doi.org/10.1007/s00190-007-0146-0

Vaidyanathan, P. (1993). *Multirate systems and filter banks*. Prentice Hall. Retrieved from https://books.google.com/books?id=pAsfAQAAIAAJ

Varenius, E., Haas, R., & Nilsson, T. (2021). Short-baseline interferometry local-tie experiments at the Onsala Space Observatory. *Journal of Geodesy, 95*(5), 54. https://doi.org/10.1007/s00190-021-01509-5

Virtanen, P., Gommers, R., Oliphant, T., Haberland, M., Reddy, T., Cournapeau, D., et al. (2020). Scipy 1.0: Fundamental algorithms for scientific computing in python. *Nature Methods, 17*(3), 1–12. https://doi.org/10.1038/s41592-019-0686-2

York, J. (2021). HRTRs at VLBA sites supporting foundation CORS and beyond. In *61st meeting of the civil GPS service interface committee*.

York, J., Little, J., Munton, D., & Barrientos, K. (2011). A fast number-theoretic transform approach to a GPS receiver. *Navigation, 57*(4), 297–307. https://doi.org/10.1002/j.2161-4296.2010.tb01784.x

York, J., Little, J., Munton, D., & Barrientos, K. (2014). A detailed analysis of GPS live-sky signals without a dish. *Navigation, 61*(4), 311–322. https://doi.org/10.1002/navi.69

York, J., Little, J., Nelson, S., Caldwell, O., & Munton D. (2012). A novel software defined GNSS receiver for performing detailed signal analysis. In *Proceedings of the 2012 international technical meeting of the institute of navigation* (pp. 1910–1921).